%% file: main.tex
\documentclass[lettersize,journal]{IEEEtran}
    \usepackage{algorithm}
    \usepackage{algpseudocode}
    \usepackage{amsmath}
    \usepackage{amssymb}
    \usepackage{multicol}
    \usepackage{multirow}
    \usepackage{orcidlink}
    \usepackage{cite}
    \usepackage{caption}
    \usepackage{subcaption}
    \usepackage{tikz}
    \usepackage{pgfplots}
    \newcommand{\circnum}[1]{\tikz[baseline=(char.base)]{\node[shape=circle,draw,inner sep=1pt] (char) {#1};}}
    \usepackage{hyperref}

    \usepackage{soul}
    \usepackage{listings}
    \usepackage[table]{xcolor}
    \pgfplotsset{compat=1.18}

    \newcommand{\malik}[1]{{\color{black}#1}}

    \definecolor{commentblue}{RGB}{0,70,140}
    \lstdefinestyle{cstyle}{
      language=C,
      basicstyle=\ttfamily\scriptsize,
      columns=fullflexible,
      keepspaces=true,
      showstringspaces=false,
      frame=single,
      breaklines=true,
      commentstyle=\color{blue}\itshape
    }

\begin{document}

\title{\textsf{QUISA}: A Compact, Flexible and Unified Instruction-Set Coprocessor for QR-UOV Signature}

\author{
        Malik~Imran~\orcidlink{0000-0002-1900-6387}, 
        Ciara~Rafferty~\orcidlink{0000-0002-3670-366X},
        Ayesha~Khalid~\orcidlink{0000-0002-4815-6966} and
        ~Máire~O'Neill~\orcidlink{0000-0002-6865-6212}
\thanks{
This work is funded by the Integrated Quantum Networks (IQN) Research  Hub (EP/Z533208/1).\textit{(Corresponding Author: Malik Imran)}
            
            All authors are with the Center for Secure Information Technologies (CSIT), Queen's University Belfast (QUB), NI, United Kingdom. (e-mail: \{m.imran@qub, c.m.rafferty@qub, a.khalid@qub, m.oneill@ecit.qub\}.ac.uk)

    
        }     
}



\maketitle

\begin{abstract}
        QR-UOV is a Quotient Ring Unbalanced Oil and Vinegar (UOV) multivariate quadratic (MQ) signature scheme that has progressed to round-3 of the post-quantum cryptography (PQC) additional signatures call by the National Institute of Standards and Technology (NIST).
        Until now, for architectural efficiency and performance trade-offs, QR-UOV has been evaluated only on software platforms and has never been considered on hardware, such as field-programmable gate arrays (FPGAs).
        To address this, we are the first to provide a complete QR-UOV hardware implementation through a flexible and unified \textsf{QUISA} coprocessor design, which supports instruction-driven execution for key generation, signing, and verification across all three NIST security levels ($\texttt{SL-I}$, $\texttt{SL-III}$, $\texttt{SL-V}$). 
        The proposed architecture features extensive throughput optimizations by parallelizing key QR-UOV module operations, such as the pseudorandom byte generator, hashing, and rejection sampling. To minimize hardware resources, we propose a modular arithmetic unit with finite-field reduction based on the shift–add folding method. The design introduces a linear system solver for \textsf{QUISA} based on LU decomposition with partial pivoting and a sample solution module, operating over the prime field $\mathbb{F}_{\mathrm{q}}$ ($\mathrm{q=127}$).
        We comprehensively evaluate \textsf{QUISA} on an Artix-7 FPGA at 80~MHz up to the post-place-and-route stage, utilizing 11,144 slices and 35,248 look-up tables. For higher \texttt{SL-V} security levels, it completes key generation, signing, and verification in 30.4M, 3.8M, and 2.8M clock cycles, respectively.

    \end{abstract}


    \section{Introduction}

    Quantum computing has advanced rapidly in recent years, driven by continued improvements in quantum processor scale and capability. The increasing availability of larger quantum processors has intensified concerns regarding the long-term security of conventional public-key cryptography.
    Fujitsu and RIKEN, together at the RIKEN RQC-Fujitsu Collaboration Center, have developed 64-qubit and 256-qubit quantum processors, with a roadmap that targets a 1,000-qubit system by 2026~\cite{fujitsu}. Similarly, IBM has reported successive generations of quantum processors, including Eagle (127 qubits), Heron (133 qubits), and Nighthawk (120 qubits), with plans to scale its architecture toward larger systems in the coming years, such as a 360-qubit processor by 2026~\cite{ibm}. 
    These advancements indicate that future quantum computers may provide substantial computational power. Meanwhile, quantum algorithms, by Shor~\cite{shors97} and Grover~\cite{Grover96}, threaten the security of widely deployed public-key cryptosystems, such as RSA~\cite{RSA1978} and elliptic curve cryptography (ECC)~\cite{ecc_by_Miller1985}.
    
    To address these risks, the National Institute of Standards and Technology (NIST) launched the post-quantum cryptography (PQC) standardization initiative in 2016~\cite{NIST_2016_Initiative}. Following several rounds of evaluation, NIST selected CRYSTALS-Kyber~\cite{Kyber} as a key encapsulation mechanism (KEM) and CRYSTALS-Dilithium~\cite{dilithium}, Falcon~\cite{falcon}, and SPHINCS+~\cite{sphincs} as digital signature schemes in 2022. These algorithms were standardized by the Federal Information Processing Standards (FIPS)~203 (ML-KEM, derived from CRYSTALS-Kyber), FIPS~204 (ML-DSA, derived from CRYSTALS-Dilithium), and FIPS~205 (SLH-DSA, derived from SPHINCS+), while the FALCON-based standard (FN-DSA/FIPS~206) is under development. In 2025, NIST also selected HQC as an alternative KEM to diversify the key encapsulation portfolio.
    To further diversify digital signature constructions, NIST issued an additional call for PQC signature schemes in 2022. From 40 complete submissions, 9 were selected for the ongoing round-3 evaluation in 2026, including the Quotient Ring Unbalanced Oil and Vinegar (QR-UOV) signature scheme~\cite{qruov_initially_proposed}.

    QR-UOV is a multivariate quadratic (MQ) signature scheme that employs a quotient-ring-based algebraic framework over a finite field to reduce the public-key size compared to the conventional Unbalanced Oil and Vinegar (UOV) constructions. 
    The original UOV scheme was introduced in~\cite{UOV_initially_proposed}, while QR-UOV was first proposed in~\cite{qruov_initially_proposed}. A refined QR-UOV design was later submitted to NIST for standardization, with its specification in~\cite{qruov_round2_spec}. To explore the trade-off between sign-key compactness and signing efficiency, selected algorithms from~\cite{qruov_round2_spec} were implemented in~\cite{qruov_software_imp} on modern x86 processors.
    More recently,~\cite{qruov_x86_processors} has accelerated QR-UOV on modern x86 architectures by exploiting arithmetic over small odd prime-power extension fields to leverage hardware multipliers and SIMD vector instructions.
    Beyond software implementations, \cite{rejscore} introduced the lightweight \texttt{RejSCore} architecture, which combines an AES-CTR-based pseudorandom number generator (\texttt{PRG}) with an iterative rejection sampling procedure to reduce hardware area at the cost of clock cycles.

    
    \subsection{Limitations of Prior SoTA}

    Among the ongoing NIST round-3 MQ-based digital signatures, only UOV~\cite{OV} and MAYO~\cite{HaMAYO, whipping_MAYO} have complete hardware implementations. In contrast, QR-UOV hardware is limited to a rejection-sampling module~\cite{rejscore}, while complete implementations have been reported only in software~\cite{qruov_software_imp, qruov_x86_processors}. To the best of the author's knowledge, no complete QR-UOV hardware architecture exists for field-programmable gate arrays (FPGAs) or application-specific integrated circuit (ASICs) platforms. Additional critical concerns are as follows: 
    
    \begin{itemize}
        \item 
        Existing accelerators~\cite{OV, HaMAYO, whipping_MAYO} target high-performance, relying on dedicated datapaths such as 2$\times$, 3$\times$, and 4$\times$ AES128 instances in~\cite{whipping_MAYO}, resulting in area-intensive designs for resource-constrained embedded devices.
    
        \item 
        Prior FPGA and ASIC implementations~\cite{OV, HaMAYO, whipping_MAYO} follow two approaches: (i) coprocessors that provide flexibility at the expense of parallelism~\cite{OV, whipping_MAYO}, and (ii) dedicated processors that maximize parallelism but support only a single security level~\cite{HaMAYO}. None of these supports the parallel operation of functional modules for key generation, signing, and verification while maintaining flexibility across multiple security levels in a single design.
    
        \item 
        Existing accelerators~\cite{OV, HaMAYO, whipping_MAYO} employ dedicated datapaths per security level, causing both hardware area and clock cycles to scale with the parameter set.
    \end{itemize}

    Consequently, the state-of-the-art (SoTA) lacks an architecture that supports the parallel operation of functional modules for key generation, signing, and verification while maintaining flexibility across multiple security levels within a fixed hardware footprint, which is a key requirement for practical deployment~\cite{sapphire, sujoy_saber}. In addition, existing linear system solvers~\cite{OV, HaMAYO, whipping_MAYO} target Gaussian elimination over the binary fields $\mathbb{F}_{2^4}$ or $\mathbb{F}_{2^8}$, whereas QR-UOV requires arithmetic over the prime field $\mathbb{F}_{127}$, which no existing MQ-based hardware solver supports.


    \subsection{Contributions}
    
    To address these gaps, our main contribution in this work is a complete QR-UOV hardware implementation through a flexible and unified \textsf{QUISA}\footnote{\textsf{QUISA} is a \underline{Q}R-\underline{U}OV-based \underline{i}nstruction-\underline{s}et coprocessor \underline{a}rchitecture. It is implemented at the register-transfer level (RTL).
    } 
    coprocessor design, supporting instruction-driven execution of key generation, signing, and verification across all NIST-defined security levels (\texttt{SL-I}, \texttt{SL-III}, \texttt{SL-V}) within a fixed hardware footprint. We summarize our additional contributions below:

    \begin{itemize}

        \item \malik{To reduce the clock cycles of the proposed \textsf{QUISA} design, we present a parallel pseudorandom sampling and packing unit (\texttt{PPSPU}) architecture that operates the SHAKE-128/256 wrapper in parallel with rejection sampling. A dual-datapath SHAKE-128/256 wrapper is implemented to support \texttt{PRG} and \texttt{Hash} operations specific to MQ-based PQC algorithms, as well as standard FIPS~202-compliant computations. For rejection sampling, an efficient hardware architecture is presented in section~\ref{subsubsec:arch_rej_samp}, based on a minor modification to the round-2 QR-UOV reference implementation.}
        
        \item To minimize the area, we propose an iterative architecture for the modular arithmetic unit of \textsf{QUISA} in section~\ref{subsec:arch_au}. Since the modulus $\mathrm{q}$ in QR-UOV is a Mersenne prime, modular reduction is realized efficiently using shift–add folding operations.

        \item We propose a linear system solver for \textsf{QUISA} comprising LU decomposition with partial pivoting and a sample solution module, operating over the prime field $\mathbb{F}_{\mathrm{127}}$. 
        
        \item \textsf{QUISA} is implemented up to the post-place-and-route (\texttt{Post-PAR}) stage and evaluated on an \texttt{Artix-7} FPGA. At 80 MHz, it utilizes 11,144 slices and 35,248 look-up tables to implement all security levels in a unified design. For the NIST highest security level of \texttt{SL-V}, it completes key generation, signing, and verification in 30M, 3M, and 2M clock cycles.
        
    \end{itemize}


    All modules in \textsf{QUISA} follow the NIST round-2 QR-UOV specification except for rejection sampling: \texttt{RejSamp()}. The \texttt{RejSamp()} in the reference C/C++ implementation comprises \texttt{rejsamp\_and\_$\mathrm{q}$()} and \texttt{rejsamp\_rejection\_with\_aux()}. The former function is implemented without changes, whereas the latter is modified in this work to operate \texttt{PRG} and rejection sampling in parallel to reduce the clock cycles. Consequently, this change modifies the reference implementation and the corresponding KAT files. Thus, we used the modified KATs to verify \textsf{QUISA}. Apart from this modification, \textsf{QUISA} remains fully compliant with the NIST round-2 reference implementation. Further details are in section~\ref{subsubsec:arch_rej_samp}.


    \section{Preliminaries}\label{sec:background}


    \subsection{Notation}\label{subsec:notations}

    Let $\mathbb{F}_\mathrm{q}$ denote the finite field of order $\mathrm{q}$, where $\mathrm{q}$ is a Mersenne prime, and let $\mathbb{F}_\mathrm{q}[x]$ denote the polynomial ring over $\mathbb{F}_\mathrm{q}$. QR-UOV operates over the quotient ring $\mathbb{F}_\mathrm{q}[x]/(f)$, where $f \in \mathbb{F}_\mathrm{q}[x]$ is an irreducible polynomial of degree $\ell$.
    Lets $\mathbb{F}_\mathrm{q}^{\ell}$ denote the $\ell$-dimensional vector space over $\mathbb{F}_\mathrm{q}$, under which each element of $\mathbb{F}_\mathrm{q}[x]/(f)$ can be represented as a vector in $\mathbb{F}_\mathrm{q}^{\ell}$. Let $\mathbb{F}_\mathrm{q}^{\ell \times \ell}$ denote the set of $\ell \times \ell$ matrices over $\mathbb{F}_\mathrm{q}$. For $g \in \mathbb{F}_\mathrm{q}[x]/(f)$, let $\Phi_g \in \mathbb{F}_\mathrm{q}^{\ell \times \ell}$ denote the matrix representation of multiplication by $g$. $\mathcal{A}_f = \{\Phi_g \mid g \in \mathbb{F}_\mathrm{q}[x]/(f)\}$ is the set of matrices corresponding to multiplication in the quotient ring.
    Let $W \in \mathbb{F}_\mathrm{q}^{\ell \times \ell}$ be a matrix such that $WX$ is symmetric for any $X \in \mathcal{A}_f$. $W\mathcal{A}_f = \{WX \mid X \in \mathcal{A}_f\}$ denotes the corresponding transformed set. $\mathcal{A}_f^{a,b}$ represents the set of $a\ell \times b\ell$ block matrices whose components belong to $\mathcal{A}_f$. $W^{(a)}$ shows the block-diagonal matrix.
    The $\ell$, $V$, and $M$ are positive integers. The number of vinegar variables is $v = \ell V$, the number of oil variables (which also equals the number of equations) is $m = \ell M$, and the total number of variables is $n = v + m$. $N = V + M$ is a positive integer. $\lambda$ is a security parameter. $r$ is a random $\lambda$-bit string. $\tau$ denotes the length of the random byte string used as input to the rejection sampling, which takes different values ($\tau_1$, $\tau_2$, $\tau_3$) depending on its usage in key generation, signing, and verification. Table~\ref{tab:sec_parameters} summarizes the QR-UOV security parameters. 


    \begin{table*}[tb]
    \centering
    \small
    \setlength{\tabcolsep}{9.5pt}
    \caption{QR-UOV: main security parameters~\cite{qruov_round2_spec}. 
    }
    \begin{tabular}{l|c|c|c|c|c|c|c|c|c|c}
    \hline

    \multirow{1}{*}{\textbf{Security}} & 
    \multirow{2}{*}{\textbf{$\mathrm{q}$, $v$, $m$, $\ell$}} & 
    \multirow{1}{*}{\textbf{Public/Private}} & 
    \multirow{1}{*}{\textbf{Signature}} &
    \multirow{2}{*}{$n$} & 
    \multirow{2}{*}{$N$} & \multirow{2}{*}{$V$} & 
    \multirow{2}{*}{$M$} & 
    \multirow{2}{*}{$\tau_1$} & 
    \multirow{2}{*}{$\tau_2$} & 
    \multirow{2}{*}{$\tau_3$} \\

    {\textbf{Level}} & {} & {\textbf{Keys (Bytes)}} & {\textbf{(Bytes)}} & {} & {} & {} & {} & {} & {} & {} \\ \hline

    {\texttt{SL-I}} & 127,156,54,3 & 24,255/32 & 200 & 210 & 70 & 52 & 18 & 4267 & 2916 & 82 \\

    {\texttt{SL-III}} & 127,228,78,3 & 71,891/48 & 292 & 306 & 102 & 76 & 26 & 9020 & 6123 & 120 \\

    {\texttt{SL-V}} & 127,306,105,3 & 173,676/64 & 392 & 411 & 137 & 102 & 35 & 16144 & 11018 & 162 \\ 

    \hline
    \end{tabular}
    \label{tab:sec_parameters}
    \end{table*}

    \subsection{QR-UOV}   

    
    QR-UOV consists of key generation (KeyGen), signing (Sign), and verification (Verify) operations, given in Algorithms~\ref{alg:kg},~\ref{alg:sign},~and~\ref{alg:verify}, where bold lowercase and italic uppercase letters denote vectors and matrices, respectively.
    In KeyGen, two random seeds are sampled. The secret seed is expanded using $\texttt{Expand}_{sk}$ to obtain the matrix $\bar{S}'$, while the public seed is used in $\texttt{Expand}_{pk}$ to derive matrices $(\bar{P}_{i,1}, \bar{P}_{i,2})$ for each $i$. These are then used to compute $\bar{P}_{i,3}$ (public quadratic components).
    In Sign, $\bar{S}'$ and $(\bar{P}_{i,1}, \bar{P}_{i,2})$ are derived through $\texttt{Expand}_{sk}$ and $\texttt{Expand}_{pk}$, respectively. Vinegar variables $\mathbf{y}$ are sampled using SHAKE. Using $(\bar{P}_{i,1}, \bar{P}_{i,2})$ for each $i$, intermediate computations yield the linear coefficients $\mathbf{L}_i$ and constants $u_i$. These form the system $L\mathbf{x} = \mathbf{t} - \mathbf{u}$, where $\mathbf{t}$ is obtained by hashing the message with randomness $r$. If a solution exists, the oil variables are derived; otherwise, the process repeats with new randomness. Finally, $\mathbf{y}$ and the oil variables, together with $\bar{S}'$, produce $\mathbf{s}$, and Sign outputs $(r,\mathbf{s})$.
    In Verify, the verifier recomputes $\mu$ using the public seed and message, and derives the target vector $\mathbf{t}$ using randomness $r$. For each $i$, $(\bar{P}_{i,1}, \bar{P}_{i,2})$ is regenerated via $\texttt{Expand}_{pk}$. The corresponding quadratic evaluation is then performed on $\mathbf{s}$ to obtain the verification value. The signature is accepted if this matches $\mathbf{t}$; otherwise, it is rejected.
    
    
    \begin{algorithm}[tb]
    \caption{KeyGen()~\cite{qruov_round2_spec}}
    \begin{algorithmic}[1]
    
    \Require Security parameter $\lambda$
    \Ensure Public key $pk \in \{0,1\}^{\lambda} \times (\mathbb{F}_{\mathrm{q}^{\ell}}^{M \times M})^{m}$ and secret key $sk \in \{0,1\}^{2\lambda}$
    
    \State $(\text{seed}_{pk}, \text{seed}_{sk}) \overset{\$}{\leftarrow} \{0,1\}^{\lambda}$
    
    \State $\bar{S}' \leftarrow \texttt{Expand}_{\texttt{sk}}(\text{seed}_{sk})$ 
    \Comment{$\bar{S}' \in \mathbb{F}_{\mathrm{q}^{\ell}}^{V \times M}$}
    
    \For{$i = 1$ to $m$}
        \State $(\bar{P}_{i,1}, \bar{P}_{i,2}) \leftarrow \texttt{Expand}_{\texttt{pk}}(\text{seed}_{pk}, i)$
        \Comment{$\bar{P}_{i,1} \in \mathbb{F}_{\mathrm{q}^{\ell}}^{V \times V}$ (symmetric), $\bar{P}_{i,2} \in \mathbb{F}_{\mathrm{q}^{\ell}}^{V \times M}$}
        
        \State $\bar{P}_{i,3} \leftarrow -\bar{S}'^{T}\bar{P}_{i,1}\bar{S}' + \bar{P}_{i,2}^{T}\bar{S}' + \bar{S}'^{T}\bar{P}_{i,2}$
        \Comment{$\bar{P}_{i,3} \in \mathbb{F}_{\mathrm{q}^{\ell}}^{M \times M}$}
    \EndFor
    
    \State $pk \leftarrow (\text{seed}_{pk}, \{\bar{P}_{i,3}\}_{i \in [m]})$, $sk \leftarrow (\text{seed}_{sk}, \text{seed}_{pk})$
    \State \Return $(pk, sk)$
    
    \end{algorithmic}
    \label{alg:kg}
    \end{algorithm}

    
    \begin{algorithm}[tb]
    \caption{Sign($M$, $sk$)~\cite{qruov_round2_spec}}
    \begin{algorithmic}[1]
    
    \Require Message $M \in \mathbb{B}^*$ and private key $sk \in \{0,1\}^{2\lambda}$
    \Ensure Signature $\sigma \in \{0,1\}^{\lambda} \times \mathbb{F}_\mathrm{q}^m$
    
    \State $(\text{seed}_{sk}, \text{seed}_{pk}) \leftarrow sk$
    
    \State $\bar{S}' \leftarrow \texttt{Expand}_{\texttt{sk}}(\texttt{seed}_{\texttt{sk}})$ \Comment{$\texttt{SD} \in \mathbb{F}_{\mathrm{q}^l}^{V \times M}$}
    
    \State $\mathbf{y} = (y_1, \dots, y_v)^T \overset{\$}{\leftarrow} \mathbb{F}_\mathrm{q}^v$ \Comment{$\mathbf{y}$ is a vector of vinegar elements ($\texttt{y}\leftarrow \texttt{Expand}_{\texttt{y}}(\texttt{seed}_{\texttt{y}})$)}
    
    \For{$i = 1$ to $m$}
        \State $(\bar{P}_{i,1}, \bar{P}_{i,2}) \leftarrow \textcolor{black}{\text{Expand}_{pk}}(\text{seed}_{pk}, i)$ 
        \State $\bar{\mathbf{y}} \leftarrow W^{(V)}\phi^{-1}(\bar{P}_{i,1})\mathbf{y}$ \Comment{$\bar{\mathbf{y}} \in \mathbb{F}_\mathrm{q}^v$}
        \State $\mathbf{L}_i \leftarrow -2\phi^{-1}(\bar{S}')^T \bar{\mathbf{y}} + 2W^{(V)}\phi^{-1}(\bar{P}_{i,2}^T)\mathbf{y}$ \Comment{$\mathbf{L}_i \in \mathbb{F}_\mathrm{q}^m$}
        \State $u_i \leftarrow \mathbf{y}^T \bar{\mathbf{y}}$ \Comment{$u_i \in \mathbb{F}_\mathrm{q}$}
    \EndFor
    
    \State $L \leftarrow (\textbf{L}_1, \dots, \textbf{L}_m)^T$ \Comment{$L \in \mathbb{F}_\mathrm{q}^{m \times m}$}
    \State $\mathbf{u} \leftarrow (u_1, \dots, u_m)^T$ \Comment{$\mathbf{u} \in \mathbb{F}_\mathrm{q}^m$}
    \State $\mu \leftarrow \textcolor{black}{\text{SHAKE-256}}(\text{seed}_{pk} \| \textcolor{black}{\text{BytesToBits}}(M), 512)$
    
    \Repeat
        \State $r \overset{\$}{\leftarrow} \{0,1\}^{\lambda}$
        \State $\textbf{t} \leftarrow \textcolor{black}{\text{Hash}}(\mu, r)$ \Comment{$\textbf{t} \in \mathbb{F}_\mathrm{q}^m$}
    \Until{$L\mathbf{x} = \textbf{t} - \mathbf{u}$ has solutions for $\mathbf{x}$}
    
    \State Choose one solution $(y_{v+1}, \dots, y_n)^T \in \mathbb{F}_\mathrm{q}^{m}$ of $L\mathbf{x} = \textbf{t} - \mathbf{u}$ randomly
    \State $\mathbf{s} \leftarrow (y_1, \dots, y_v, y_{v+1}, \dots, y_n)^T - (\phi^{-1}(\bar{S}')\cdot (y_1, \dots, y_v)^T \| \textbf{0}_m)$ \Comment{$\mathbf{s} \in \mathbb{F}_\mathrm{q}^n$}
    \State \Return $\sigma = (r, \mathbf{s})$
    \end{algorithmic}
    \label{alg:sign}
    \end{algorithm}

    
    \begin{algorithm}[!htbp]
    \caption{Verify($M$, $pk$, $\sigma$)~\cite{qruov_round2_spec}}
    \begin{algorithmic}[1]
    
    \Require Message $M \in \mathbb{B}^*$, $pk \in \{0,1\}^{\lambda} \times \left( \mathbb{F}_\mathrm{q}^{M \times M} \right)^m$ and $\sigma \in \{0,1\}^{\lambda} \times \mathbb{F}_\mathrm{q}^m$
    \Ensure Accept or reject
    
    \State $(\text{seed}_{pk}, \{\bar{P}_{i,3}\}_{i \in [m]}) \leftarrow pk$
    \State $(r, \mathbf{s}) \leftarrow \sigma$
    
    \For{$i = 1$ to $m$}
        \State $(\bar{P}_{i,1}, \bar{P}_{i,2}) \leftarrow \textcolor{black}{\text{Expand}_{pk}}(\text{seed}_{pk}, i)$
        \State $P_i \leftarrow W^{(N)} \phi^{-1} 
        \begin{pmatrix}
        \bar{P}_{i,1} & \bar{P}_{i,2} \\
        \bar{P}_{i,2}^\top & \bar{P}_{i,3}
        \end{pmatrix}
        $ \Comment{$P_i \in \mathbb{F}_\mathrm{q}^{n \times n}$}
    \EndFor
    
    \State $\mu \leftarrow \textcolor{black}{\text{SHAKE-256}}(\text{seed}_{pk} \| \textcolor{black}{\text{BytesToBits}}(M), 512)$
    \State $\mathbf{t} \leftarrow \textcolor{black}{\text{Hash}}(\mu, r)$
    \State $\mathbf{t'} \leftarrow \left( \mathbf{s}^T P_1\textbf{s}, \dots, \mathbf{s}^T P_m\textbf{s} \right)^T$ \Comment{$\mathbf{t'} \in \mathbb{F}_\mathrm{q}^m$}
    \State \Return accept if $\mathbf{t} = \mathbf{t'}$, and reject otherwise
    
    \end{algorithmic}
    \label{alg:verify}
    \end{algorithm}

    Algorithms~\ref{alg:kg},~\ref{alg:sign},~and~\ref{alg:verify} require several auxiliary functions, as described below. For additional details, we refer readers to~\cite{qruov_round2_spec}. 
    

    \begin{itemize}
        \item $\texttt{Expand}_{sk}$ expands $\text{seed}_{sk}$ into matrix $\bar{S}' \in \mathbb{F}_{\mathrm{q}^\ell}^{V \times M}$. Let $n_2 = \ell V M$. A sequence $(v_1,\dots,v_{n_2})$ is generated using $\texttt{RejSampPRG}(\text{seed}_{sk}, 1,\tau_2, n_{2})$ and assigned to $\bar{S}'$.

        \item $\texttt{Expand}_{pk}$ expands $\text{seed}_{pk}$ with index $i$ to generate $(\bar{P}_{i,1}, \bar{P}_{i,2})$. First, $n_1 = \ell V (V+1)/2$ elements are generated via $\texttt{RejSampPRG}(\text{seed}_{pk}, 2i-1, \tau_1, n_{1})$ and mapped to $\bar{P}_{i,1}$ using $\texttt{ExpandSymmetricMatrixVxV()}$. Then, $n_2 = \ell V M$ elements are generated via $\texttt{RejSampPRG}(\text{seed}_{pk}, 2i, \tau_2, n_{2})$ and stored in $\bar{P}_{i,2}$ using $\texttt{ExpandMatrixVxM()}$. 

        \item $\texttt{Hash}$ computes the value $\mathbf{t} \in \mathbb{F}_\mathrm{q}^m$ from the message-dependent 64-bit value $\mu$ and the salt $r$. A byte sequence $(r_1,\dots,r_{\tau_3})$ is generated by $\texttt{SHAKE\text{-}256}(\texttt{BytesToBits}(\mu)\|r,8\tau_{3})$, followed by $\texttt{RejSamp}(r_1,\dots,r_{\tau_3}, \tau_3, m)$ to obtain $\mathbf{t}$. 

        \item $\texttt{RejSampPRG}$ generates a pseudorandom sequence $(v_1,\dots,v_{n'}) \in \mathbb{F}_\mathrm{q}^{n'}$ from $(\text{seed}, i, \tau, n')$. It comprises the following two steps: 

            \begin{itemize}
                \item First, $(r_1,\dots,r_{\tau})$ is produced via $\texttt{PRG}(\text{seed},i,8\tau)$. \texttt{PRG} can be performed using AES-CTR or SHAKE: for the AES option, 128, 192, and 256-bit variants are required for \texttt{SL-I}, \texttt{SL-III}, and \texttt{SL-V}, respectively, while for the SHAKE option, SHAKE-128 is used for \texttt{SL-I} and SHAKE-256 for \texttt{SL-III} and \texttt{SL-V}.
                
                \item Second, $\texttt{RejSamp}((r_1,\dots,r_{\tau}), \tau, n')$ is applied to obtain valid field elements of $(v_1,\dots,v_{n'})$. It is rejection sampling.
                
            \end{itemize}

    \end{itemize}

    The main and auxiliary functions, including $\texttt{Expand}_{sk}$, $\texttt{Expand}_{pk}$, $\texttt{Hash}$, $\texttt{SHAKE-256}$, $\texttt{PRG}$, $\texttt{RejSampPRG}$, and $\texttt{RejSamp}$ in Algorithms~\ref{alg:kg}--\ref{alg:verify}, are implemented by the \texttt{PPSPU} (section~\ref{subsec:arch_ppspu}). The remaining functions, including $\texttt{ExpandSymmetricMatrixVxV()}$, $\texttt{ExpandMatrixVxM()}$, and transpose operations, rely on efficient memory scheduling instead of dedicated hardware.


    \section{\textsf{QUISA} Coprocessor Architecture} \label{sec:quisa}



    \subsection{\malik{Host-Controlled High-Level Interface}}  
    \label{subsec:host_controller}

\begin{figure*}[htbp]
    \centering
    \includegraphics[width=0.85\textwidth]{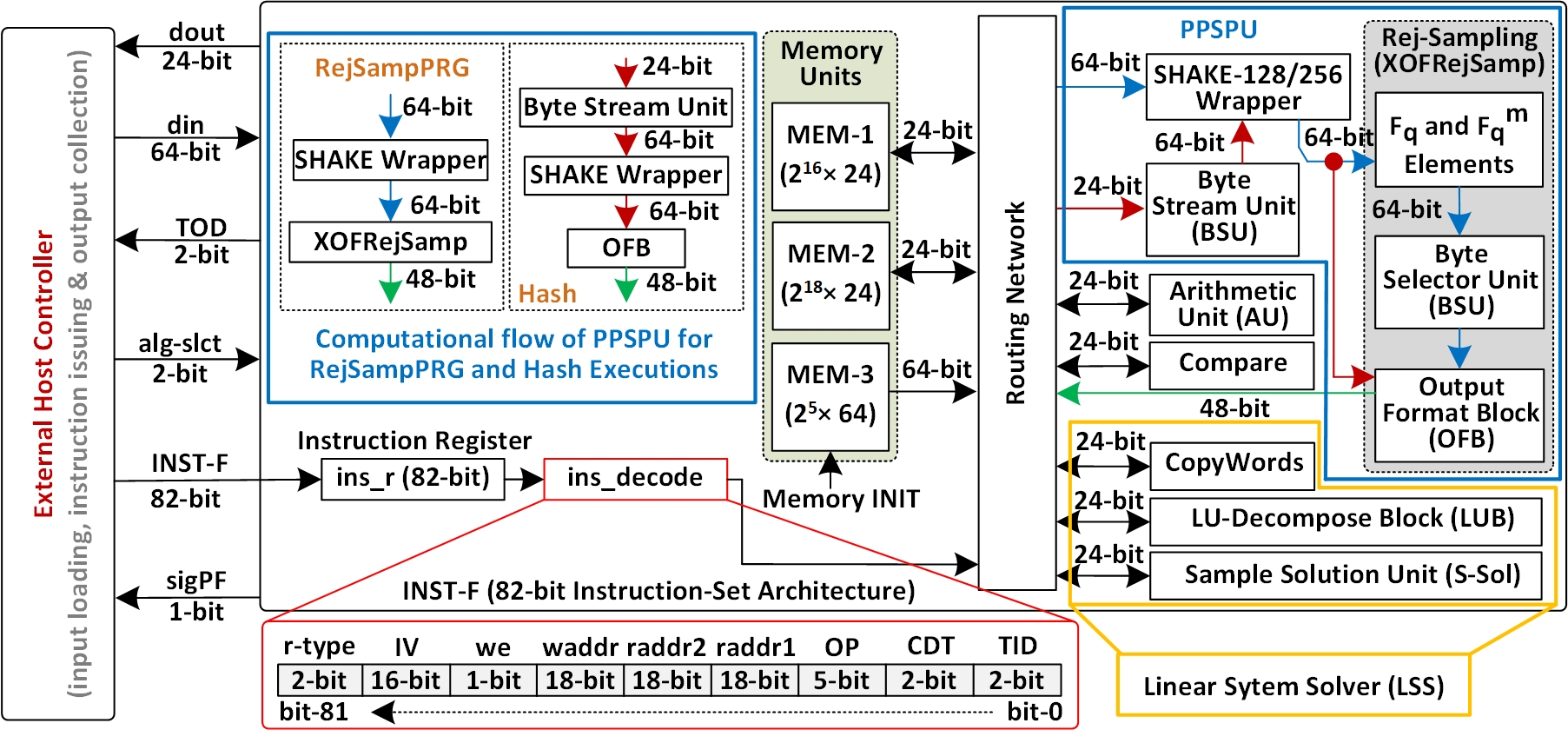}
    \caption{\malik{Host-controlled \textsf{QUISA} coprocessor architecture.}}
    \label{fig:quisa}
    \end{figure*}

    \malik{
    Fig.~\ref{fig:quisa} shows the host-controlled \textsf{QUISA} coprocessor, which supports KeyGen, Sign, and Verify across all three NIST security levels using a unified configurable datapath. The external host communicates with \textsf{QUISA} through the 64-bit \texttt{din}, 24-bit \texttt{dout}, and 2-bit \texttt{TOD} interface, where \texttt{TOD} indicates instruction completion. The \texttt{alg-slct} selects the target operation, while \texttt{sigPF}, generated by the Compare module, indicates the signature verification result.
    To provide flexible control, \textsf{QUISA} employs an 82-bit instruction format (\texttt{INST-F}), shown in Fig.~\ref{fig:quisa}. Each instruction is issued by the host controller, stored in the instruction register (\texttt{ins\_r}), and decoded to control the routing network, memory units, and functional modules. The \texttt{TID} field specifies memory initialization, computation, or data retrieval; \texttt{CDT} identifies the input data type; \texttt{OP} defines up to 32 opcodes; \texttt{raddr1}, \texttt{raddr2}, and \texttt{waddr} specify memory base addresses; \texttt{we} enables memory writes; \texttt{IV} provides the 16-bit initialization vector for SHAKE; and \texttt{r-type} selects the SHA3/SHAKE variant.
    \textsf{QUISA} does not use an internal program memory. Instead, the external host issues one instruction at a time and waits for \texttt{TOD} before issuing the next. Table~\ref{tab:inst_storage} summarizes the instruction sequences and corresponding host-side storage for all QR-UOV operations and security levels. Supporting all three security levels requires 431 instructions, occupying 4418~B of host-side storage. For RTL verification, the testbench emulates the host by issuing these sequences, whereas in practice they may reside in ROM, flash, or other external memory. This host-controlled sequencing enables \textsf{QUISA} to interface with different platforms without modifying the coprocessor datapath.
    The decoded \texttt{INST-F} fields configure the routing network, steering data between the memory units and functional modules. While the external host provides the base read/write addresses, each functional module generates the corresponding local read/write offsets, whose sum forms the effective memory address. This instruction-driven routing enables a unified hardware implementation of KeyGen, Sign, and Verify. The memory units and functional modules are described in sections~\ref{subsec:memory_units}--\ref{subsec:system_solver}. 
    }


    \begin{table}[tb]
    \centering
    \small
    \setlength{\tabcolsep}{0.8pt}
    \caption{\textsf{QUISA} instructions. Values denote instruction count $N_{\mathrm{inst}}$ and instruction memory size ($\lceil82N_{\mathrm{inst}}/8\rceil$ bytes).}
    \label{tab:inst_storage}
    \begin{tabular}{l|c|c|c|c|c}
    \hline
    \textbf{Security} &
    \multicolumn{3}{c|}{\textbf{Per Operation Level}} &
    \textbf{Per} &
    \textbf{Unified} \\
    \cline{2-4}
    \textbf{Level}
    &
    \textbf{KeyGen}
    &
    \textbf{Sign}
    &
    \textbf{Verify}
    &
    \textbf{Security Level}
    &
    \textbf{QUISA} \\
    \hline
    
    \texttt{SL-I}
    &
    25 (257 B)
    &
    69 (708 B)
    &
    51 (523 B)
    &
    145 (1487 B)
    &
    \multirow{1}{*}{431}
    \\
    
    \texttt{SL-III}
    &
    25 (257 B)
    &
    67 (687 B)
    &
    49 (503 B)
    &
    141 (1446 B)
    &
    (4418 B) \\
    
    \texttt{SL-V}
    &
    25 (257 B)
    &
    69 (708 B)
    &
    51 (523 B)
    &
    145 (1487 B)
    &
    \\
    
    \hline
    \end{tabular}
    \end{table}


    \subsection{Memory/Storage Unit Requirements for QR-UOV} \label{subsec:memory_units}



    In QR-UOV, arithmetic is performed over an $l$-dimensional vector space defined over $\mathbb{F}_\mathrm{q}$. Following the QR-UOV specification~\cite{qruov_round2_spec}, \textsf{QUISA} packs three vector components ($l=3$) per memory address to reduce storage overhead. As shown in Fig.~\ref{fig:quisa}, the architecture employs three memory units: \texttt{MEM-1}, \texttt{MEM-2}, and \texttt{MEM-3}. \texttt{MEM-1} and \texttt{MEM-2} provide a 24-bit datapath by storing three bytes per address for QR-UOV arithmetic, whereas \texttt{MEM-3} stores 64-bit words containing the seeds and input data for the SHAKE-128/256-based \texttt{PRG}. Further details are given below:

    \begin{itemize}
        
        \item \malik{\texttt{MEM-1} stores results for KeyGen, Sign, and Verify, excluding the element-wise addition of the $V\times M$ matrix and the generation of $\bar{P}_{i,3}$. Its size is determined by the fixed address-generation schedule of the implemented algorithms, which depends only on the QR-UOV security parameters of Table~\ref{tab:sec_parameters}, and is confirmed by RTL simulation. The maximum accessed addresses are 4513 and 2586 for the Sign and Verify operations of \texttt{SL-I}, fitting within a $2^{13}$ address space. Likewise, \texttt{SL-III} and \texttt{SL-V} require $2^{14}$ and $2^{16}$ address spaces, respectively. Since each address stores three bytes, the resulting memory sizes are $2^{13}\!\times\!24$, $2^{14}\!\times\!24$, and $2^{16}\!\times\!24$ bits.
        }
        
        \item \malik{\texttt{MEM-2} stores the intermediate results of (i) the element-wise addition of the $V\times M$ matrix and (ii) the generated $\bar{P}_{i,3}$ matrix. Its size is determined directly from the QR-UOV algorithm and depends only on the parameters of Table~\ref{tab:sec_parameters}. The element-wise addition requires $V\times M$ addresses, corresponding to 936, 1976, and 3570 locations for \texttt{SL-I}, \texttt{SL-III}, and \texttt{SL-V}, respectively, whereas storing $\bar{P}_{i,3}$ requires $m(M\times M)$ addresses, resulting in 17\,496, 52\,728, and 128\,625 locations. Consequently, the total storage fits within $2^{15}$, $2^{16}$, and $2^{18}$ address spaces. Since each address stores three bytes, the resulting memory sizes are $2^{15}\!\times\!24$, $2^{16}\!\times\!24$, and $2^{18}\!\times\!24$ bits.
        }
        
        \item \malik{\texttt{MEM-3} stores the inputs to the SHAKE-128/256 wrapper, including $\text{seed}_{sk}$, $\text{seed}_{pk}$, $\text{seed}_{y}$, and $\text{seed}_{r}$. During Sign and Verify, $\text{seed}_{pk}\|\texttt{BytesToBits}(M)$ is stored in consecutive memory locations for SHAKE-128/256 to generate $\mu$. QR-UOV supports arbitrary-length messages, the KATs used for verification employ 33-byte messages. 
        Under this configuration, the largest \texttt{SL-V} input occupies 25 addresses on the 64-bit datapath, making a $2^5\times64$ \texttt{MEM-3} sufficient for all security levels. Longer messages only require increasing the depth of \texttt{MEM-3}.
        }
        
    \end{itemize}


    In \textsf{QUISA}, \texttt{MEM-1} and \texttt{MEM-2} are implemented as true dual-port BRAM IPs generated using the Vivado Block Memory Generator, whereas \texttt{MEM-3} is inferred as a simple dual-port BRAM using the \texttt{ram\_style="block"} directive. The overall memory cost depends on the target deployment scenario as discussed below.
    \circnum{1} \textbf{\textsf{QUISA} coprocessor:} Supporting all QR-UOV security levels within a unified coprocessor requires dimensioning the memory units according to the largest parameter set (\texttt{SL-V}), resulting in $2^{16}\times24$ (\texttt{MEM-1}), $2^{18}\times24$ (\texttt{MEM-2}), and $2^{5}\times64$ (\texttt{MEM-3}). The memory footprint is dominated by \texttt{MEM-2}, which stores the $\bar{P}_{i,3}$ matrices generated during on-the-fly KeyGen.
    \circnum{2} \textbf{Dedicated implementation:} When targeting a single QR-UOV security level, the memory units are dimensioned according to the selected parameter set, eliminating the over-provisioning of the unified architecture. Consequently, \texttt{SL-I} requires \texttt{MEM-1}=$2^{13}\times24$, \texttt{MEM-2}=$2^{15}\times24$, and \texttt{MEM-3}=$2^{5}\times64$. Similarly, \texttt{SL-III} requires \texttt{MEM-1}=$2^{14}\times24$, \texttt{MEM-2}=$2^{16}\times24$, and \texttt{MEM-3}=$2^{5}\times64$, while \texttt{SL-V} remains unchanged.
    \circnum{3} \textbf{Dedicated Sign implementation:} The hardware Sign engine neither generates nor stores the $\bar{P}_{i,3}$ matrices. Thus, \texttt{MEM-2} only stores the intermediate results of the $V\times M$ matrix operations, requiring $2^{10}\times24$, $2^{11}\times24$, and $2^{12}\times24$ for \texttt{SL-I}, \texttt{SL-III}, and \texttt{SL-V}, respectively, while \texttt{MEM-1} and \texttt{MEM-3} remain unchanged.
    \circnum{4} \textbf{Dedicated Verify implementation:} Unlike the hardware Sign engine, Verify requires the $\bar{P}_{i,3}$ matrices, which form part of the public key. In this work, they are preloaded into \texttt{MEM-2}, requiring $2^{15}\times24$, $2^{16}\times24$, and $2^{18}\times24$ for \texttt{SL-I}, \texttt{SL-III}, and \texttt{SL-V}, respectively. Alternatively, the $\bar{P}_{i,3}$ matrices can be streamed from external memory, reducing the on-chip \texttt{MEM-2} requirement at the expense of additional memory-transfer cycles during signature verification. This approach trades on-chip BRAM for external-memory bandwidth.


    \subsection{Parallel Pseudorandom Sampling and Packing Unit (\texttt{PPSPU})} \label{subsec:arch_ppspu}

    The blue-dotted region in Fig.~\ref{fig:quisa} forms the \texttt{PPSPU}, comprising a SHAKE-128/256 wrapper, a byte stream unit, a SHAKE-based rejection sampling unit (\texttt{XOFRejSamp}), a byte selector unit, and an output format block. All components are implemented as separate RTL modules except for the byte selector, whose functionality is integrated into \texttt{XOFRejSamp}. We refer readers to sections~\ref{subsubsec:arch_shake_wrapper}--\ref{subsubsec:ppspu_timeline} for details.


    \subsubsection{SHAKE-128/256 Wrapper + Byte Stream Unit}
    \label{subsubsec:arch_shake_wrapper} 

    The SHAKE-128/256 wrapper in \textsf{QUISA} supports the \texttt{Expand-related} functions and the \texttt{Hash} operation required in Algorithms~\ref{alg:kg}--\ref{alg:verify}. Except \texttt{Hash}, the SHAKE-128/256 wrapper loads 64-bit words directly from \texttt{MEM-3} through the routing network of Fig.~\ref{fig:quisa}. In contrast, the \texttt{Hash} operation needs to compute over the QR-UOV arithmetic data, which is stored in 24-bit words in \texttt{MEM-1} and \texttt{MEM-2}. To bridge this mismatch, a lightweight byte stream unit reads 24-bit words from two BRAM ports in parallel, extracts up to six bytes per cycle, and accumulates them into a 168-byte local buffer. Once the required bytes are collected, the buffer data is repacked into 64-bit words and forwarded to the SHAKE-128/256 wrapper for hash computations. The SHAKE-128/256 wrapper architecture is described below.
    
    \malik{The \texttt{INST-F} of \textsf{QUISA} uses a 2-bit \texttt{r-type} field to drive the SHAKE-128/256 wrapper, where, based on encoding \texttt{0} to \texttt{3}, one variant among SHA3-256, SHA3-512, SHAKE-128, and SHAKE-256 is selected. All variants share \texttt{Absorb} and \texttt{Squeeze} phases of the \texttt{KECCAK} permutation. The \texttt{Absorb} phase performs 24 rounds, each executing the $\theta$, $\rho$, $\pi$, $\chi$, and $\iota$ transformations, while the \texttt{Squeeze} phase outputs 64-bit words. \textsf{QUISA} employs a high-speed open-source \texttt{KECCAK} core~\cite{keccak_team}, where the \texttt{Absorb} and \texttt{Squeeze} phases require 24 and 21 cycles, respectively, the latter corresponding to the SHAKE-128 rate ($168/8=21$).
    }

    \malik{To interface with the \texttt{Absorb} phase of the \texttt{KECCAK} core~\cite{keccak_team}, the SHAKE-128/256 wrapper uses a 1-bit control signal, \texttt{disable\_iv\_injection}, to select the input format. When enabled, all SHA3-256, SHA3-512, SHAKE-128, and SHAKE-256 operations follow the FIPS~202 standard~\cite{fips202}. When disabled, only SHAKE-128/256 append a 2-byte \texttt{IV} before domain separation, as specified by QR-UOV~\cite{qruov_round2_spec}. This control modifies only the input formatting, leaving the \texttt{KECCAK} permutation and \texttt{Squeeze} phases unchanged at 24 and 21 cycles, respectively. Consequently, the wrapper supports both QR-UOV modes: \texttt{PRG} with \texttt{IV} insertion and \texttt{Hash} with standard SHAKE-128/256 processing.
    }

    Accordingly, our SHAKE-128/256 wrapper supports two datapaths. The blue path in Fig.~\ref{fig:quisa} corresponds to \texttt{PRG}, where a 64-bit input to the SHAKE-128/256 wrapper produces a 64-bit output that is forwarded to the \texttt{XOFRejSamp} for rejection sampling; the output formatter (green path) then writes two 24-bit words to \texttt{MEM-1} or \texttt{MEM-2} through a routing network. In contrast, the red path corresponds to \texttt{Hash}, where 24-bit inputs are processed by the byte stream unit to form a 64-bit input for the SHAKE-128/256 wrapper, and the resulting output is forwarded directly, with the formatter selecting two 24-bit words before rejection sampling for memory write-back. Thus, the SHAKE-128/256 wrapper supports QR-UOV-specific input handling while preserving standard-compliant operation, enabling its use in both QR-UOV and other PQC schemes, such as ML-KEM~\cite{fips203}, and ML-DSA~\cite{fips204}, without design modifications or additional hardware costs. 


    \subsubsection{Rejection Sampling + Byte Selector Unit + Output Format Block}
    \label{subsubsec:arch_rej_samp}

    \malik{Section~\ref{subsubsec:arch_shake_wrapper} described the SHAKE-based \texttt{PRG} in \textsf{QUISA}, which generates a pseudorandom byte stream. Each byte is ANDed with $\mathrm{q}$, producing a candidate in $\{0,\ldots,\mathrm{q}\}$. Since $\mathrm{q}$ is a Mersenne prime in QR-UOV~\cite{qruov_round2_spec}, it is the only value outside $\mathbb{F}_\mathrm{q}$ and is therefore rejected whenever it appears. Rejection sampling continues until the required number of valid field elements is obtained.
    In the round-2 reference implementation~\cite{qruov_ref_code}, the pseudorandom stream is divided into main and auxiliary regions. During rejection sampling, invalid candidates ($=\mathrm{q}$) in the main region are replaced with valid candidates from the auxiliary region, as illustrated in the left part of Fig.~\ref{fig:rejsamp_code}. Implementing this replacement mechanism in hardware requires simultaneous access to both regions, which results in irregular memory accesses and sequential dependencies that limit parallel execution of the \texttt{PRG} and rejection sampling. To overcome this limitation, \textsf{QUISA} minimally modifies the second stage of the rejection sampler, as shown in the right part of Fig.~\ref{fig:rejsamp_code}. Instead of replacing invalid main candidates with auxiliary candidates, \textsf{QUISA} sequentially compacts valid candidates from the main region, using the auxiliary region only when additional valid samples are required to reach the target output length. This modification enables regular memory accesses and allows parallel execution of both \texttt{PRG} and rejection sampling.}

\begin{figure}[tb]
\centering  
\begin{minipage}[t]{0.48\textwidth}
\centering
\begin{lstlisting}[style=cstyle]
static inline void rejsamp_rejection_with_aux(
    const unsigned int length,
    const uint8_t *aux,
    uint8_t *dst
){
    while (*aux == QRUOV_q) aux++;
    for (int i = 0; i < length; i++) {
        if (*dst == QRUOV_q) {
            *dst = *aux++;
            while (*aux == QRUOV_q) aux++;
        }
        *dst++;
    }
} // Reference: QR-UOV round-2 source code 
\end{lstlisting}
\end{minipage}
\hfill
\vspace{-2.0mm}
\begin{minipage}[t]{0.48\textwidth}
\centering
\begin{lstlisting}[style=cstyle]
static inline void rejsamp_rejection_with_aux(
    const unsigned int length,
    const uint8_t *aux,
    uint8_t *dst
){
    const unsigned int tau = length * 2;
    unsigned int j = 0;
    for (int i = 0; i < length + tau && j < length; i++) {
        uint8_t val = dst[i];
        if (val < QRUOV_q)
        dst[j++] = val;
    }
} // Modified for QUISA
\end{lstlisting}
\end{minipage}
\vspace{-2.5mm}
\caption{Reference and modified implementations of \texttt{rejsamp\_rejection\_with\_aux()}.}
\label{fig:rejsamp_code}
\end{figure}

    \malik{To implement rejection sampling in hardware, as in~\cite{rejscore}, selecting or rejecting one byte per cycle is straightforward and minimizes hardware resources, albeit at the cost of higher clock cycles. A high-performance alternative is to process multiple bytes per cycle. Although the SHAKE-based \texttt{PRG} naturally outputs 8-byte words per cycle, selecting and packing 8 valid bytes is inefficient because QR-UOV operates on $l$-dimensional vectors with $l=3$, leading to unaligned memory writes and additional read-modify-write operations. Therefore, \textsf{QUISA} selects and packs up to 3 valid bytes per cycle, naturally matching the 3-byte memory organization and reducing clock cycles compared with the one-byte-per-cycle approach of~\cite{rejscore}.
    }

\begin{figure}[tb]
    \centering
    \includegraphics[width=\columnwidth]{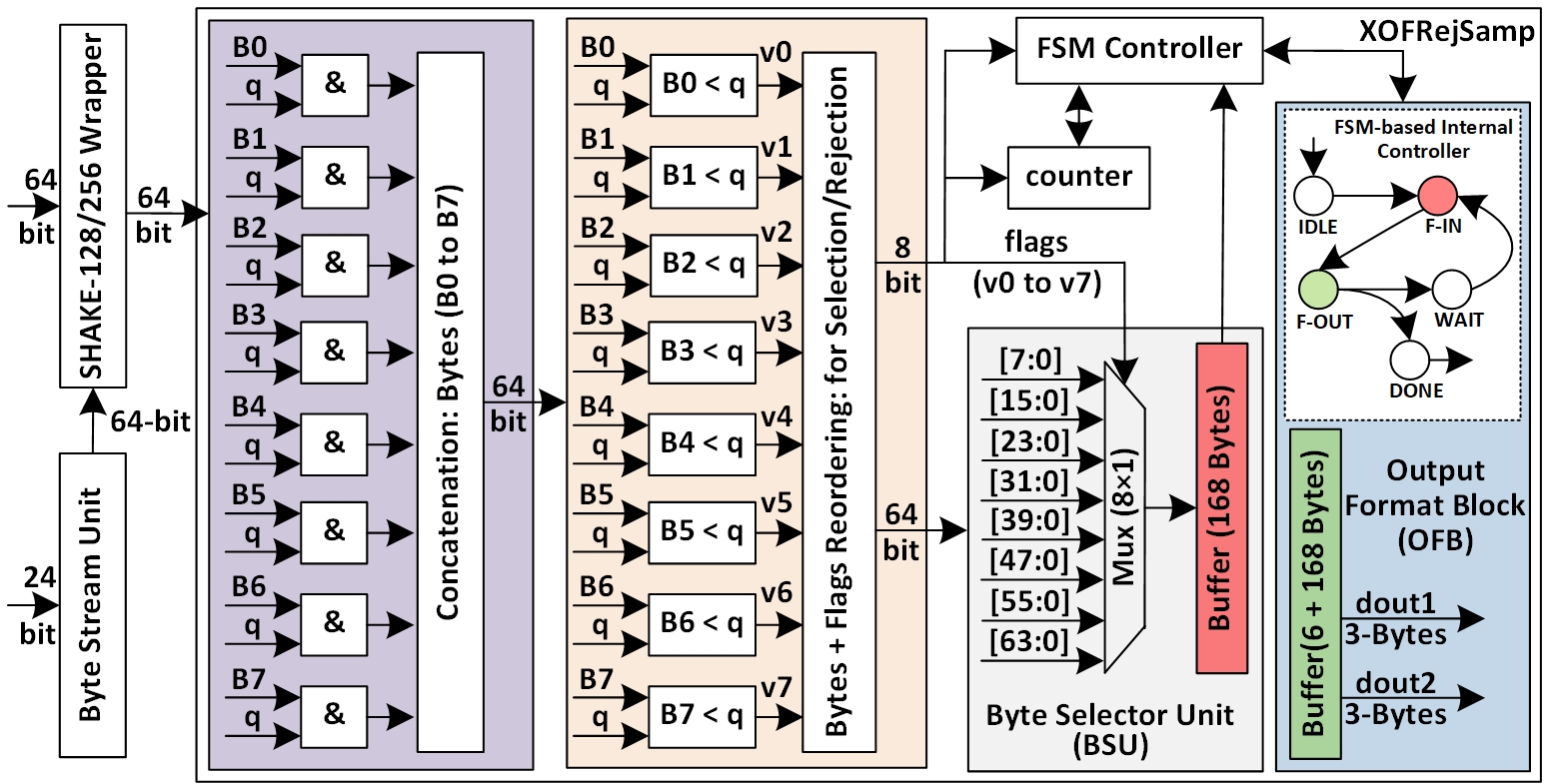}
     \caption{
    \texttt{XOFRejSamp}: SHAKE-based rejection sampling.
    }
    \label{fig:rejsampler}
    \end{figure}

    \malik{Fig.~\ref{fig:rejsampler} shows the proposed SHAKE-based rejection sampling architecture, denoted as \texttt{XOFRejSamp}. The purple block (left) implements \texttt{rejsamp\_and\_q()} by receiving 8 bytes from the SHAKE output every clock cycle and bitwise ANDing them with $\mathrm{q}$ in parallel, generating eight candidates in $\{0,\ldots,\mathrm{q}\}$ that are concatenated from \texttt{LSB} to \texttt{MSB} into a 64-bit word. The light-brown block (right) implements the rejection stage (\texttt{rejsamp\_rejection\_with\_aux()}) by comparing all eight candidates with $\mathrm{q}$ in parallel to generate the valid signals (\texttt{v0}--\texttt{v7}). Valid candidates ($<\mathrm{q}$) are packed from the \texttt{LSB} side, while non-valid candidates are shifted toward the \texttt{MSB} together with their reordered valid signals. This packing is realized using a simple combinational \texttt{if-else} circuit, allowing rejection sampling to sustain the 8-byte-per-cycle throughput of the SHAKE-based \texttt{PRG}.
    }
    

    The benefit of reordering the valid bytes and their flags is realized in the byte selector unit, where an $8\times1$ multiplexer (in Fig.~\ref{fig:rejsampler}) accumulates up to eight valid bytes per cycle, rather than one byte over eight cycles, into the red-colored buffer sized to the SHAKE-128 rate. Once all valid bytes from a 168-byte block are buffered, they are written to memory. Since the SHAKE-128/256 wrapper produces one 168-byte block every 21 cycles (8 bytes/cycle), whereas \textsf{QUISA} stores data as two 3-byte words (6 bytes/cycle), writing the entire block requires $168/6=28$ cycles. Consequently, an additional 7 cycles are needed before the next SHAKE block arrives, preventing full overlap between the SHAKE-128/256 wrapper and \texttt{XOFRejSamp}.

    The output format block in Fig.~\ref{fig:rejsampler} enables parallel execution using a lightweight controller and an additional $(168+6)$-byte buffer, where 168 bytes correspond to the SHAKE-128 rate and 6 bytes store leftover data. In the \texttt{F-IN} state, the controller stores the 168-byte output of the byte selector into the buffer in one cycle. In the \texttt{F-OUT} state, the buffer shifts by 48 bits per cycle, outputting six valid bytes through \texttt{dout1} and \texttt{dout2}. Thus, processing a $(168+6)$-byte buffer requires $6+168/6$ cycles. Complete 6-byte groups are written to \texttt{MEM-1}/\texttt{MEM-2}, while any remaining bytes (up to six) are retained as leftover data. The controller then enters the \texttt{WAIT} state for the next 168-byte SHAKE block.

    Leftover bytes occur when the number of valid bytes in the green buffer of Fig.~\ref{fig:rejsampler} is not divisible by 6. In this case, up to six valid bytes from the current \texttt{Squeeze} are retained and merged with the 168-byte output of the next \texttt{Squeeze} to preserve the QR-UOV data format. In \textsf{QUISA}, this merge is implemented using an \texttt{if--else} structure (not shown in Fig.~\ref{fig:rejsampler}). A counter tracks the valid-byte count (1--6) from the reordered valid flags during each \texttt{Squeeze} phase and passes it to the output format controller, which aligns the output of the next \texttt{Squeeze}. This process repeats until all required bytes are generated for the \texttt{Expand}-related functions or the message is fully processed for \texttt{Hash}.


    \subsubsection{Processing Timeline of the \texttt{PPSPU}} \label{subsubsec:ppspu_timeline}

\begin{figure}[tb]
        \centering
        \includegraphics[width=\columnwidth]{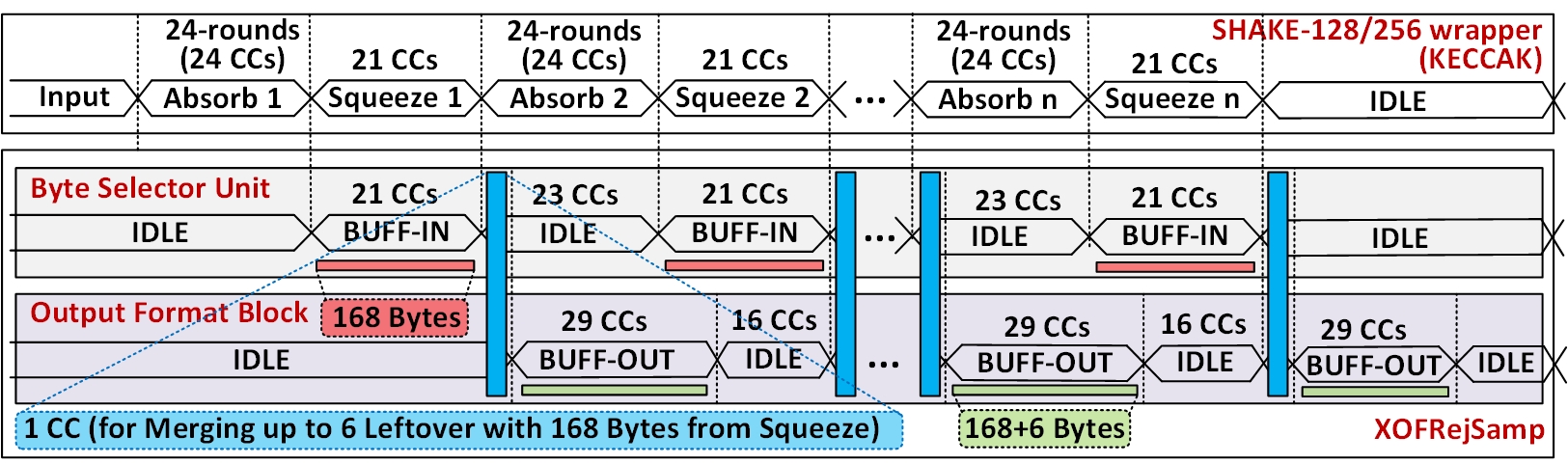}
        \caption{\texttt{PPSPU}: Parallel execution of the SHAKE-128/256 wrapper and \texttt{XOFRejSamp}.}
        \label{fig:shake_sampler_execution}
    \end{figure}

    Fig.~\ref{fig:shake_sampler_execution} shows the parallel execution of the SHAKE-128/256 wrapper and \texttt{XOFRejSamp}. The \texttt{KECCAK} core alternates between \texttt{Absorb} and \texttt{Squeeze} phases, while \texttt{XOFRejSamp} simultaneously processes the previous 168-byte output block through \texttt{BUFF-IN} and \texttt{BUFF-OUT} buffers. 
    During each 21-cycle \texttt{Squeeze} phase, \texttt{BUFF-IN} accumulates up to 168 bytes. Immediately afterward, these bytes are merged with up to 6 leftover bytes from the previous iteration into \texttt{BUFF-OUT} in a single clock cycle (blue blocks). Thereafter, \texttt{BUFF-OUT} overlaps with the subsequent \texttt{Absorb} and \texttt{Squeeze} phases, outputting 6 bytes per cycle for memory write-back over 29 cycles. This pipelined organization enables concurrent execution of the \texttt{PRG}, \texttt{Hash}, and \texttt{RejSamp} functions.
    For \texttt{SL-I} parameters with $l=3$, $V=52$, $M=18$, and ${\tau}_{2}=2916$, the AES-CTR-based rejection sampler in~\cite{rejscore} requires 8525 clock cycles to generate ${\tau}_{2}$ pseudorandom bytes and extract $lVM=2808$ valid bytes, whereas the proposed \texttt{PPSPU} completes the same operation in only 895 cycles, achieving a $9.52\times$ speedup.

    \subsection{Arithmetic Unit (AU)} \label{subsec:arch_au}

\begin{figure}[tb]
    \centering
    \includegraphics[width=\columnwidth]{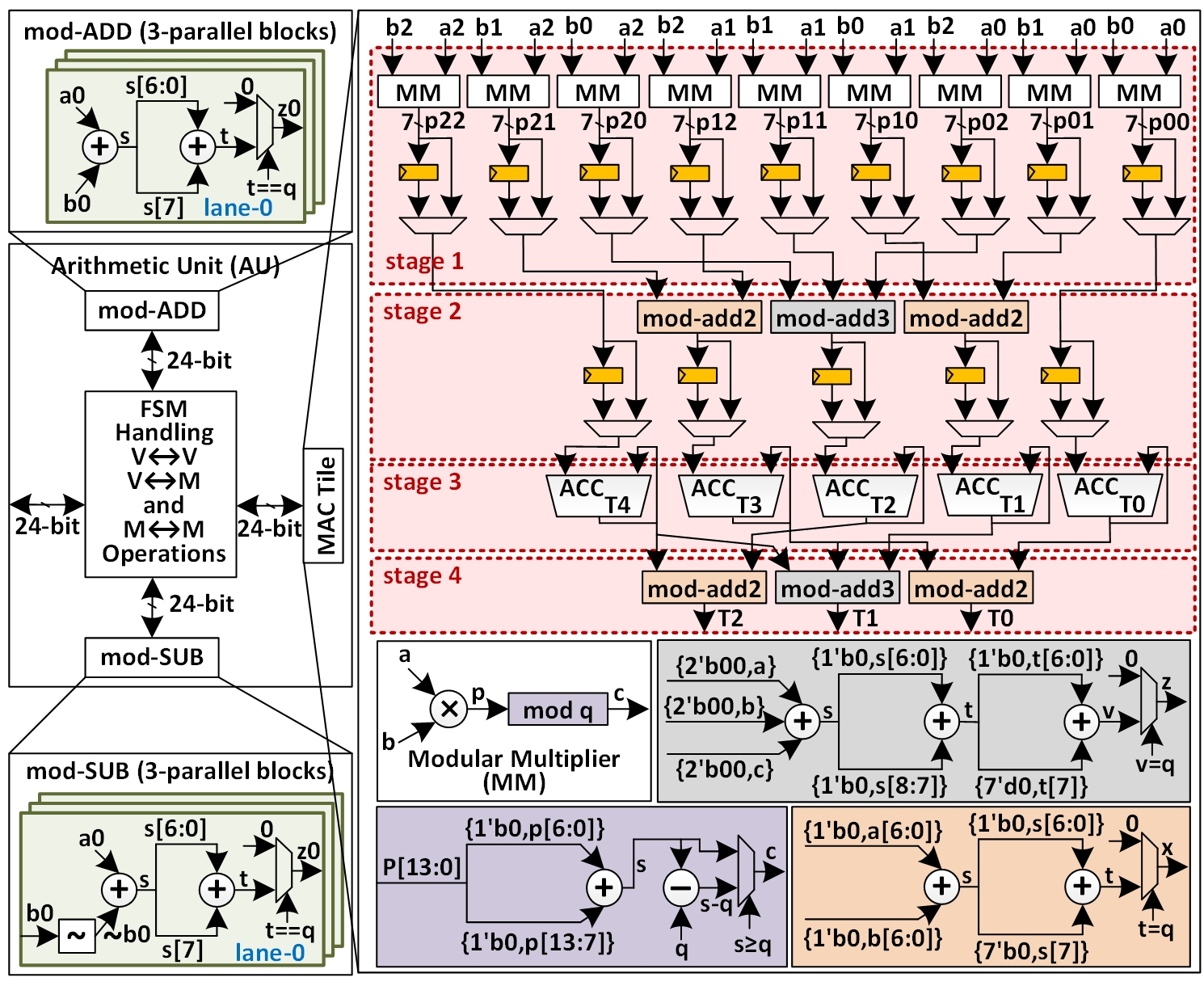}
       \caption{Arithmetic unit architecture. 
    }
    \label{fig:au}
    \end{figure}

    The arithmetic unit (AU) in \textsf{QUISA} implements the finite-field operations required by QR-UOV over $\mathbb{F}_{\mathrm{q}}^{l}$, where $\mathrm{q}=127=2^{7}-1$ is a Mersenne prime and $l=3$. As shown in Fig.~\ref{fig:au}, the AU consists of four main components: \texttt{mod-ADD}, \texttt{mod-SUB}, a MAC tile, and an FSM controller.

    
    \subsubsection{\texttt{mod-ADD} and \texttt{mod-SUB} Architectures in $\mathbb{F}_{\mathrm{q}}^{l}$} \label{subsubsec:arch_add_sub}

    The \texttt{mod-ADD} and \texttt{mod-SUB} units in Fig.~\ref{fig:au} perform three modular additions or subtractions in parallel using lanes 0--2. Each lane operates on one coefficient from the packed 24-bit inputs $(a_2,a_1,a_0)$ and $(b_2,b_1,b_0)$ to produce the corresponding coefficient of the packed output $(z_2,z_1,z_0)$. Taking lane-0 as an example, $a_0$ and $b_0$ are first combined using an 8-bit adder to generate an intermediate sum $s$. Modular reduction is then performed by an end-around-carry operation, where the carry bit is folded back into the lower seven bits as $t=s[6:0]+s[7]$. This exploits the identity $2^7 \equiv 1 \pmod{127}$, enabling an efficient reduction without dedicated modular arithmetic. A final correction stage, implemented using a $2\times1$ multiplexer, ensures canonical residues by mapping $t=\mathrm{q}$ to zero; otherwise, $t$ is forwarded as the valid result. For \texttt{mod-ADD}, the operands are directly added, i.e., $s=a_i+b_i$, whereas \texttt{mod-SUB} performs subtraction by inverting the second operand and computing $s=a_i+(\sim b_i)$. The same datapath is replicated across all three lanes, allowing three modular results to be produced in a single clock cycle.


    \subsubsection{Multiplication in $\mathbb{F}_{\mathrm{q}}^{l}$} \label{subsubsec:march_mul}

    Let $\mathbf{a}=(a_2,a_1,a_0)$ and $\mathbf{b}=(b_2,b_1,b_0)$ denote two vector elements in $\mathbb{F}_{\mathrm{q}}^{3}$. Their product $\mathbf{z}=(z_2,z_1,z_0)$ is computed as $z_0=a_0b_0+a_1b_2+a_2b_1,\;
    z_1=a_0b_1+a_1b_0+a_2b_2,\;
    z_2=a_0b_2+a_1b_1+a_2b_0
    $, where all operations are performed modulo $\mathrm{q}$. This computation requires nine coefficient multiplications and six modular additions, with each output coefficient formed by the sum of three partial products. A conventional hardware implementation would accumulate these partial products using cascaded two-input modular adders. In contrast, the proposed MAC tile in Fig.~\ref{fig:au} employs both \texttt{mod-add2} (two-input) and \texttt{mod-add3} (three-input) units, allowing three partial products to be accumulated in a single stage. This reduces the addition depth and shortens the critical path of the MAC datapath.
    
    
    \paragraph{MAC Tile} 

    It is organized into four stages: (i) modular multiplication, (ii) a sum tree, (iii) accumulation, and (iv) final recombination, as described below:  
    
    To initiate the processing, nine modular multipliers (MMs) compute the pairwise products $p_{ij} = a_i b_j$, where $i, j \in \{0,1,2\}$, each operating on 7-bit operands to produce a 14-bit product reduced modulo $\mathrm{q}$ via the Mersenne identity $2^{7} \equiv 1 \pmod{127}$ through a fold and conditional subtraction (purple blocks in Fig.~\ref{fig:au}). Since such small-width multiplications do not efficiently utilize DSP slices, each MM is implemented using a \texttt{*} operator in RTL, allowing the synthesis tool to use FPGA LUTs efficiently. The reduced products are registered in orange boxes in Fig.~\ref{fig:au} using $2\times1$ multiplexers. These multiplexers allow users the choice to configure the MAC with different pipeline depths. 
    In the second stage, the nine reduced products are organized into a sum tree of two \texttt{mod-add2} units and one \texttt{mod-add3} unit, computing $sT_1 = p_{01} + p_{10}$, $sT_2 = p_{02} + p_{11} + p_{20}$, and $sT_3 = p_{12} + p_{21}$ in a single clock cycle, as these adders are fully combinational. The outputs of $sT_1$--$sT_3$ together with $p_{00}$ and $p_{22}$ (which bypass the sum tree) are routed to their respective accumulators through $2\times1$ multiplexers (see Fig.~\ref{fig:au}). These multiplexers provide the same pipeline depth flexibility as stage 1.
    In the third stage, as shown in Fig.~\ref{fig:au}, the five accumulator registers $\texttt{ACC}_{T0}$--$\texttt{ACC}_{T4}$ accumulate $p_{00}$, $sT_1$, $sT_2$, $sT_3$, and $p_{22}$, respectively, retaining partial sums over $n$ cycles, where $n$ is the length of the input vector pair being processed (either $V$ or $M$ words depending on the operation). One input word pair is consumed per cycle, allowing the tile to operate as a true multiply-accumulate unit over vectors of arbitrary length.
    In the final stage, the accumulated values are recombined in a single cycle through two \texttt{mod-add2} and one \texttt{mod-add3} units, as shown in Fig.~\ref{fig:au}, to produce the three output coefficients: 
    $T_0 = \texttt{ACC}_{T0} + \texttt{ACC}_{T3}, \quad
    T_1 = \texttt{ACC}_{T1} + \texttt{ACC}_{T3} + \texttt{ACC}_{T4}, \quad
    T_2 = \texttt{ACC}_{T2} + \texttt{ACC}_{T4}$, which correspond to $z_0$, $z_1$, and $z_2$ respectively, and are packed into a 24-bit word for write-back to memory.

   The MAC tile supports a configurable pipeline depth via \texttt{MAC\_STAGES}, which automatically inserts optional register stages after the MM outputs and the sum tree to improve timing closure. Fig.~\ref{fig:au} shows the general MAC architecture with both optional register stages. However, the implementation results in section~\ref{sec:results_and_comparisons} correspond to \texttt{MAC\_STAGES}$=2$, where only the register stage after the MM outputs is active. With \texttt{MAC\_STAGES}$=2$, the MAC pipeline introduces \texttt{MAC\_LAT}$=1$ cycles of latency. Since the BRAM read latency is 2 cycles, the MAC tile internally manages both latencies through a shift register (\texttt{mac\_cons\_sr}) that delays the input valid signal by \texttt{MAC\_LAT} cycles to assert the accumulation enable exactly when valid data arrives from memory. This gives a total latency of $n+2$ cycles for an inner product of length $n$.


    \subsubsection{Address Scheduling for mod-ADD, mod-SUB and MAC Tile} \label{subsubsec:fsm_au}

    In Fig.~\ref{fig:au}, the \texttt{mod-ADD}, \texttt{mod-SUB}, and the MAC tile compute operations over a vector-element. The FSM controller generates the read and write addresses and controls the execution of \texttt{mod-ADD}, \texttt{mod-SUB}, and MAC tile operations according to the QR-UOV parameters $V$ and $M$. For \texttt{mod-ADD} and \texttt{mod-SUB}, two 24-bit operands are read sequentially from memory units, and the resulting vector element is written back. In particular, for \texttt{mod-ADD}, \texttt{MEM-1} and \texttt{MEM-2} operate concurrently: one provides the input operands, while the other stores the result in the same cycle, reducing latency. The \texttt{mod-SUB} also executes sequentially, where operands are read, the subtraction is performed, and the result is written back to the same memory before the next operands are accessed, introducing a small clock cycle overhead. For both \texttt{mod-ADD} and \texttt{mod-SUB}, the controller generates linear addresses of length $V$ or $M$, supporting vector–vector (V$\leftrightarrow$V), vector–matrix (V$\leftrightarrow$M), and matrix–matrix (M$\leftrightarrow$M) operations to be executed iteratively.

    Similar to \texttt{mod-ADD} and \texttt{mod-SUB}, the MAC tile supports V$\leftrightarrow$V, V$\leftrightarrow$M, and M$\leftrightarrow$M operations iteratively. For these operations, the FSM generates three sequences of read/write addresses: linear (increment by 1), stride-based (increment by $V$ or $M$), and triangular (index-dependent offset with decreasing stride). Linear addressing accesses consecutive elements, while stride-based addressing uses a fixed offset to access matrix columns stored in row-major order. The benefit of stride-based addressing is evident in operations such as $\bar{P}_{i,2}^{T}\bar{S}'$ and $\bar{S}'^{T}\bar{P}_{i,2}$ in Algorithm~\ref{alg:kg}, which require accessing the transpose of $\bar{P}_{i,2}$ and $\bar{S}'$. Since these matrices are stored in row-major order, their transposed elements are accessed using a stride equal to the matrix dimension ($V$ or $M$), eliminating the need to store explicit transposed matrices and reducing both memory usage and access overhead.

    For symmetric matrix operations, the controller employs triangular address generation in two scenarios. First, during symmetric matrix construction, only the non-redundant upper triangular elements are written into memory. Second, during symmetric matrix multiplication, only the non-redundant terms are fetched for processing. In both cases, the first operand address increments linearly, while the second operand address is initialized with an index-dependent offset and updated with a progressively decreasing stride to avoid redundant memory accesses. For all addressing modes (linear, stride-based, and triangular), the MAC tile consumes one operand pair per cycle and accumulates the corresponding products. Similar to \texttt{mod-SUB}, the MAC tile operates sequentially: operands are read, the inner product is accumulated, and the result is written back before the next operand pair is accessed. Thus, the same arithmetic datapath supports symmetric matrix construction, symmetric matrix multiplication, and the general V$\leftrightarrow$V, V$\leftrightarrow$M, and M$\leftrightarrow$M operations required by QR-UOV.

    The iterative read/write address scheduling, using linear, stride-based, and triangular access patterns for \texttt{mod-ADD}, \texttt{mod-SUB}, and MAC tile computations, allows the arithmetic unit to reuse the same datapath of Fig.~\ref{fig:au} across KeyGen, Sign, and Verify. Also, this strategy reduces hardware resource utilization with a modest increase in clock cycles.


    \subsection{Linear System Solver (\texttt{LSS})} \label{subsec:system_solver}

\begin{figure}[htbp]
    \centering
    \includegraphics[width=\columnwidth]{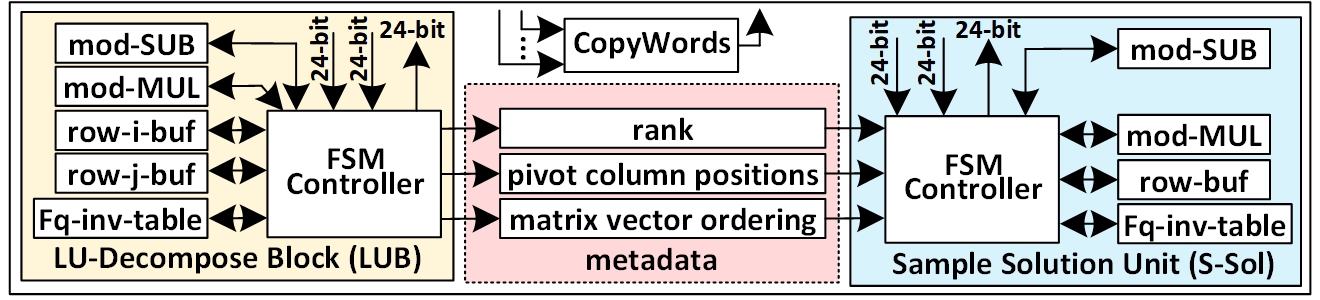}
    \caption{Linear system solver: operating \texttt{LUB}, \texttt{CopyWords} \& \texttt{S-Sol} modules independently.
    }
    \label{fig:lss}
    \end{figure}

    During signing, the linear system solver (\texttt{LSS}) solves $L\mathbf{x}=\mathbf{t}-\mathbf{u}$ over $\mathbb{F}_{127}$, where $L$ is constructed from the vinegar vector $\mathbf{y}$ (Algorithm~\ref{alg:sign}, lines~4--10), $\mathbf{x}$ denotes the oil variables, and $\mathbf{t}-\mathbf{u}$ is the target vector. Since $L$ is independent of the random salt $r$, LU decomposition with partial pivoting is performed once before the repeat loop (lines~13--16), resolving zero pivots through on-the-fly row swaps. The probability that $L\mathbf{x}=\mathbf{t}-\mathbf{u}$ has no solution is at most $1/\mathrm{q}\approx0.8\%$~\cite{qruov_round2_spec}. Thus, the specification suggests choosing a new salt $r$ until at least one solution is obtained. 
    Note that since \textsf{QUISA} takes a sequence of instructions from the host controller, choosing another salt $r$ requires only modifying the sequence of instructions rather than making any changes in the datapath of the proposed architecture.
    This entire procedure is realized in hardware using three modules: (i) an LU decomposition block (\texttt{LUB}), (ii) a sample solution unit (\texttt{S-Sol}), and (iii) a CopyWords unit, as shown in Fig.~\ref{fig:lss}.  
    The CopyWords unit is used for two purposes. First, the vector elements $\mu_i$ generated in line~8 of Algorithm~\ref{alg:sign} contain only a single valid byte at the LSB position of each 24-bit memory word. After that, CopyWords reads three consecutive words, extracts their valid LSB bytes, packs them into a single valid 24-bit word, and writes the result back to memory. This process continues until all valid bytes are packed. Second, CopyWords supports direct word-level copying, which is reused to store the generated signature contiguously in memory. The architectures of \texttt{LUB} and \texttt{S-Sol} are described in sections~\ref{subsubsec:arch_lu_decompose} and~\ref{subsubsec:arch_sample_sol}, respectively.

    
    \subsubsection{LU-Decompose Block (\texttt{LUB})} \label{subsubsec:arch_lu_decompose}

    The \texttt{LUB} module (left panel of Fig.~\ref{fig:lss}) reads two 24-bit words per cycle and produces one 24-bit word for write-back. It also generates metadata comprising the rank, pivot-column positions, and matrix vector ordering after swapping, which is stored and updated on-the-fly in the \textsf{QUISA} top module. This metadata enables the sample solution unit to reconstruct the solution according to the pivot structure and original vector ordering. The \texttt{LUB} performs four FSM-controlled operations: (i) matrix vector loading, (ii) pivot search with vector swapping, (iii) normalization, and (iv) elimination. Here, a matrix vector denotes a sequence of $M$ words.

    The \texttt{LUB} module operates on two matrix vectors: the pivot vector $i$ and the current vector $j$, streamed into the local buffers \texttt{row\_i\_buf} and \texttt{row\_j\_buf}. Each stores $M$ words corresponding to $\ell \times M$ coefficients over $\mathbb{F}_{127}$. Loading a vector requires $M+2$ cycles, including two BRAM-latency cycles. A new pivot requires loading both vectors; otherwise, only vector $j$ is reloaded while the pivot vector remains buffered.

    Once both vectors are buffered, the controller checks whether the pivot element of \texttt{row\_i\_buf} at column $c$ is non-zero; otherwise, rows $j$ are loaded into \texttt{row\_j\_buf} iteratively until a non-zero entry is found, or $c$ is advanced if none exists. This continues until a valid pivot is found or all $m$ columns are exhausted, handling rank-deficient matrices by recording the echelon rank and completing the decomposition without stalling or restarting. The two buffers are then swapped, written back to memory in $M$ cycles, and \texttt{orig\_row\_id} is updated to preserve the original row ordering.

    When a valid pivot is identified, the pivot-column index and rank metadata are updated. The pivot vector is then normalized in-place within \texttt{row\_i\_buf} using the multiplicative inverse of the pivot element, retrieved from a precomputed ROM-based lookup table (\texttt{Fq\_inv\_table}) over $\mathbb{F}_{127}$. The remaining elements are scaled by the inverse using a dedicated combinational multiplier with Mersenne reduction. This multiplier is separate from the AU MAC tile to avoid scheduling overhead. Since one field element is processed per cycle, normalization requires at most $m-c-1$ cycles, where $c$ is the current pivot-column index. The normalized vector is written back to memory and simultaneously retained in \texttt{row\_i\_buf} for immediate reuse during elimination.

    The elimination is applied to each vector $j$ below the pivot vector $i$. The elimination factor $\mu$ is the element at pivot column $c$ of \texttt{row\_j\_buf}, read in a single cycle. The vector is then updated as $\mathbf{r}_j \leftarrow \mathbf{r}_j - \mu\,\mathbf{r}_i$, where modular multiplication and subtraction are performed using lightweight combinational circuits, processing one field element per cycle. Updating the remaining $m - c - 1$ elements therefore requires at most $m - c - 1$ cycles. The updated vector is repacked into 24-bit words and written back to memory in $M$ cycles. This procedure repeats for all vectors $j > i$ to complete a single elimination pass.

    \textsf{QUISA} repeats the above steps until LU decomposition is complete or all pivot columns are exhausted. The metadata is tracked on-the-fly in dedicated buffers and becomes immediately available to \texttt{S-Sol} (section~\ref{subsubsec:arch_sample_sol}).
    

    \subsubsection{Sample Solution Unit (S-Sol)} \label{subsubsec:arch_sample_sol}

    Like \texttt{LUB}, \texttt{S-Sol} is used only during Sign to solve the linear system. It reads 24-bit words from memory to load the LU-factored matrix vectors of $L$ and the elements of $\mathbf{b}=\mathbf{t}-\mathbf{u}$, and writes the solution $\mathbf{x}$ back upon completion. \texttt{S-Sol} also receives the metadata generated by \texttt{LUB} to reconstruct $\mathbf{x}$. This process consists of three FSM-controlled phases: (i) matrix vector loading, (ii) forward substitution, and (iii) backward substitution.

    A vector of the LU-factored matrix $L$ is streamed from memory into a local buffer \texttt{row\_buf}, storing $M$ words corresponding to $\ell \times M$ field coefficients, requiring $M+2$ cycles including two cycles of BRAM latency. The metadata received from \texttt{LUB} is unpacked combinationally into two local arrays: \texttt{orig\_row\_id}, which retains the original row ordering prior to swapping and is used during forward substitution to index into $\mathbf{b}$ correctly, and \texttt{index\_map}, which retains the pivot-column position of each pivot row and is used during backward substitution to place each solution element at its correct position in $\mathbf{x}$.

    Once a vector of $L$ is loaded into \texttt{row\_buf}, \texttt{S-Sol} computes the intermediate vector $\mathbf{b}_2$ by performing forward substitution over the $r$ pivot rows, where $r$ is the rank provided by \texttt{LUB} metadata. For each pivot row $i$, the dot product $\sum_{k=0}^{i-1} L[i][k] \cdot \mathbf{b}_2[k]$ is accumulated over $i$ cycles into \texttt{T\_ACC} using a 21-bit accumulator with Mersenne reduction. The corresponding element $\mathbf{b}[\texttt{orig\_row\_id}[i]]$ is then read from memory in three cycles, including two cycles of BRAM latency, and $\mathbf{b}_2[i]$ is computed as $\mathbf{b}_2[i] \leftarrow \left( \mathbf{b}[\texttt{orig\_row\_id}[i]] - \texttt{T\_ACC} \right) \cdot L[i][i]^{-1}$, where $L[i][i]^{-1}$ is retrieved from \texttt{Fq\_inv\_table} over $\mathbb{F}_{127}$ and modular multiplication and subtraction are performed using lightweight combinational circuits. The computed $\mathbf{b}_2[i]$ is stored in \texttt{b2\_buf} and simultaneously written back to memory, packed into 24-bit words of three field coefficients.

    When forward substitution is complete, \texttt{S-Sol} recovers $\mathbf{x}$ by performing backward substitution in reverse order over the $r$ pivot rows. Any free variable not associated with a pivot column is assigned zero directly into \texttt{x\_buf} via a decrementing pointer \texttt{i\_ptr}, which advances until it reaches the pivot-column position \texttt{index\_map}$[j]$ of the current pivot row $j$. For each pivot row $j$, traversed from $j = r-1$ down to $j = 0$, the dot product $\sum_{k=\texttt{index\_map}[j]+1}^{m-1} L[j][k] \cdot \mathbf{x}[k]$ is accumulated over $m - \texttt{index\_map}[j] - 1$ cycles into \texttt{T\_ACC}, and the solution element is then computed as $\mathbf{x}[\texttt{index\_map}[j]] \leftarrow \mathbf{b}_2[j] - \texttt{T\_ACC}$ using a lightweight combinational subtraction circuit, and stored in \texttt{x\_buf} at position \texttt{index\_map}$[j]$. Once all pivot rows are processed, the complete solution vector $\mathbf{x}$ is streamed from \texttt{x\_buf} to memory over $M$ cycles, with three field coefficients packed into each 24-bit word.

    Overall, \texttt{S-Sol} reconstructs $\mathbf{x}$ from the LU-factored matrix $L$, the vector $\mathbf{b}$, and the metadata produced on-the-fly by \texttt{LUB}, completing the linear system solution over $\mathbb{F}_{127}$ in a hardware realization tightly coupled to the decomposition block.

  
    \section{Implementation Results and Comparisons}\label{sec:results_and_comparisons}

    The \textsf{QUISA} coprocessor is manually modeled in \texttt{Verilog-HDL} using \texttt{Vivado IDE} and implemented up to the \texttt{Post-PAR} stage on the \texttt{Artix-7 (xc7a200tsbv484-3)} FPGA device. The \texttt{Post-PAR} execution encompassed KAT-based validation of all functional blocks, i.e., KeyGen, Sign, and Verify operations vs. all three security levels and benchmarked against the \malik{round-2} reference implementation by the QR-UOV team~\cite{qruov_ref_code}. 
    

    
    \subsection{\malik{Clock-Cycle Counts and }Timing Results} \label{subsec:res_timing}

    \malik{
    Table~\ref{tab:cc_time} reports the clock cycles, execution time, and throughput for the KeyGen, Sign, and Verify operations. The clock cycles are obtained from the behavioral simulation of the full instruction sequence. The execution time is calculated as $\frac{\text{Clock Cycles}}{f_{\text{clk}}}$ at $f_{\text{clk}} = 80$~MHz, and the throughput is derived as $\frac{1000}{\text{Time (ms)}}$ in operations per second (ops/s).    
    The clock cycles, computation time, and throughput columns in Table~\ref{tab:cc_time} reveal that KeyGen is the computational bottleneck at all security levels due to the large public key, which grows $7.2\times$ from \texttt{SL-I} to \texttt{SL-V}, while the corresponding cycle count grows by only $5.7\times$, as it scales with the matrix dimensions $(v, m)$ rather than the total public key size. In contrast, Sign and Verify are significantly more efficient: at \texttt{SL-I}, they require only $10\%$ and $7.5\%$ of the KeyGen cycle count, respectively, reflecting the compact signature sizes of QR-UOV. For all security levels, Verify outperforms Sign by $1.3\times$--$1.4\times$ in clock cycles.
    }   
    
    
    \begin{table*}[htbp]
        \centering
        \caption{Clock cycles, execution time, and throughput @ 80~MHz.}
        \label{tab:cc_time}
        \small
        \setlength{\tabcolsep}{9.5pt}
        \begin{tabular}{l|c c c| c c c| c c c}
        \hline
        \multirow{1}{*}{\textbf{Security}} & 
        \multicolumn{3}{c|}{\textbf{Clock Cycles}} &
        \multicolumn{3}{c|}{\textbf{Time (ms)}} &
        \multicolumn{3}{c}{\textbf{Throughput (ops/s)}} \\ \cline{2-10}
        
        {\textbf{Levels}} & 
        {\textbf{KeyGen}} & 
        {\textbf{Sign}} & 
        {\textbf{Verify}} & 
        {\textbf{KeyGen}} & 
        {\textbf{Sign}} & 
        {\textbf{Verify}} &
        {\textbf{KeyGen}} & 
        {\textbf{Sign}} & 
        {\textbf{Verify}} \\ \hline
        
        \texttt{SL-I}   & 5,330,536  & 535,697   & 399,777   & 66.63  & 6.69  & 4.99  & 15.01 & 149.48 & 200.40 \\
        \texttt{SL-III} & 22,594,572 & 1,607,428 & 1,194,832 & 282.43 & 20.09 & 14.93 & 3.54  & 49.78  & 66.98  \\
        \texttt{SL-V}   & 30,415,770 & 3,809,819 & 2,801,059 & 380.19 & 47.62 & 35.01 & 2.63  & 20.99  & 28.56  \\
        \hline
        \end{tabular}
    \end{table*}

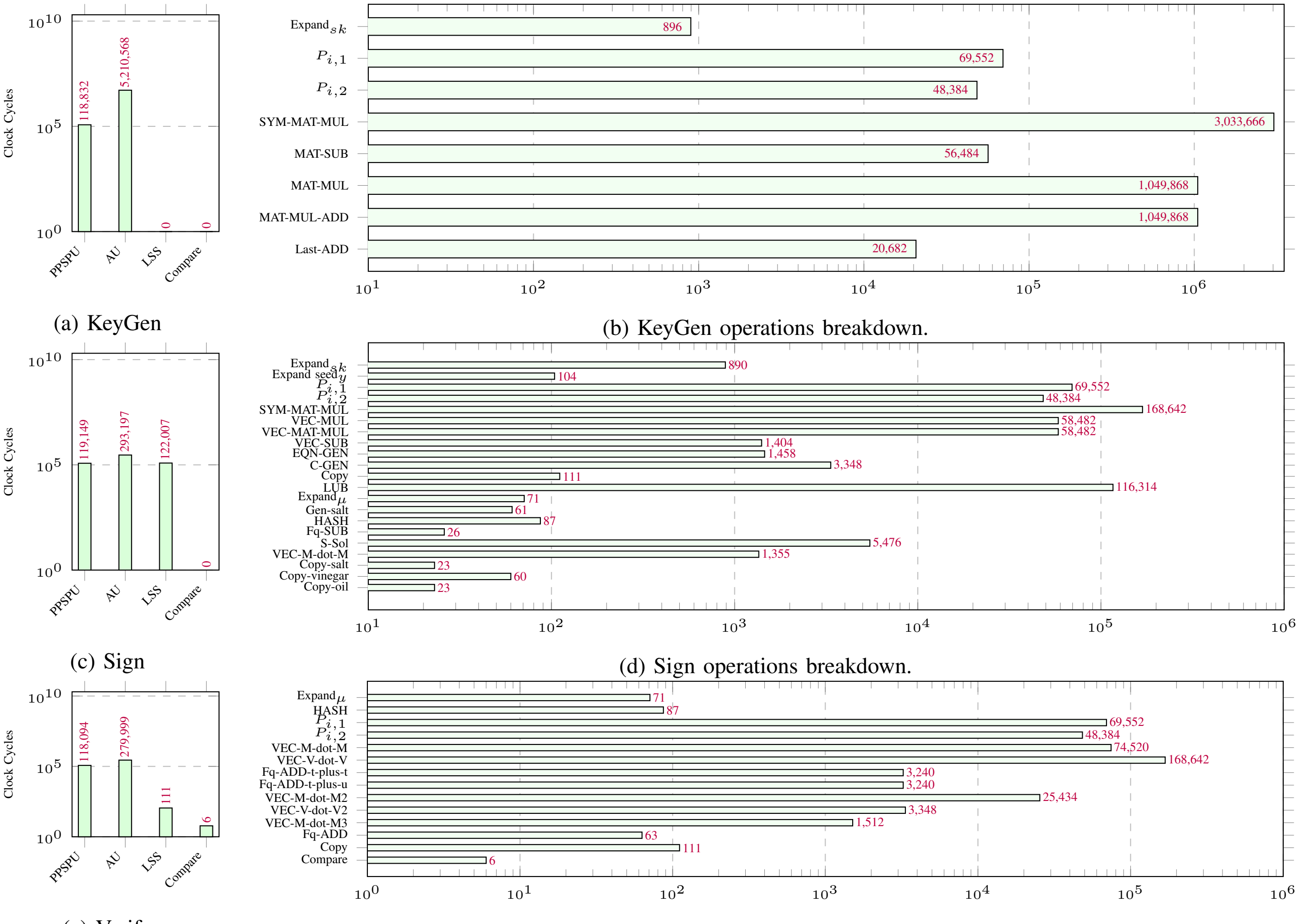
\begin{figure*}
        \centering
        \begin{subfigure}[t]{0.20\textwidth}
        \centering
        \begin{tikzpicture}[baseline=(current bounding box.north)]
        \begin{axis}[
            ybar,
            bar width=5pt,
            width=\linewidth,
            height=4.6cm,
            ymin=1,
            ymax=2e10,
            ymode=log,
            log basis y=10,
            enlarge x limits={abs=5pt},
            ylabel={Clock Cycles},
            ylabel style={
                font=\tiny,
                at={(axis description cs:-0.325,0.5)},
                anchor=south
            },
            yticklabel style={font=\tiny},
            symbolic x coords={PPSPU,AU,LSS,Compare},
            xtick=data,
            x tick label style={
                font=\tiny,
                rotate=45,
                anchor=east
            },
            ymajorgrids=true,
            grid style=dashed,
            nodes near coords,
            every node near coord/.append style={
                text=purple,
                rotate=90,
                anchor=west,
                inner sep=1pt,
                font=\tiny
            },
            point meta=explicit symbolic,
            clip=false,
        ]
        \addplot[
            fill=green!15,
            draw=black,
            line width=0.4pt
        ] coordinates {
            (PPSPU,118832) [118,832]
            (AU,5210568)   [5,210,568]
            (LSS,1)        [0]
            (Compare,1)    [0]
        };
        \end{axis}
        \end{tikzpicture}
        \caption{KeyGen}
        \label{fig:keygen_modules}
        \end{subfigure}
        \hfill
        \begin{subfigure}[t]{0.79\textwidth}
        \centering
        \begin{tikzpicture}[baseline=(current bounding box.north)]
        \begin{axis}[
            xbar,
            bar width=7pt,
            xmode=log,
            log basis x=10,
            xmin=10,
            xmax=3500000,
            width=\linewidth,
            height=5.3cm,
            x tick label style={font=\tiny},
            yticklabel style={font=\tiny},
            symbolic y coords={
                Last-ADD,
                MAT-MUL-ADD,
                MAT-MUL,
                MAT-SUB,
                SYM-MAT-MUL,
                $P_{i,2}$,
                $P_{i,1}$,
                Expand$_{sk}$
            },
            ytick=data,
            point meta=explicit symbolic,
            nodes near coords={\pgfplotspointmeta},
            every node near coord/.append style={
                font=\tiny,
                text=purple,
                anchor=east
            },
            xmajorgrids=true,
            grid style=dashed,
        ]
        \addplot[
            fill=green!5,
            draw=black,
            line width=0.4pt
        ] coordinates {
            (896,Expand$_{sk}$)          [896]
            (69552,$P_{i,1}$)            [69,552]
            (48384,$P_{i,2}$)            [48,384]
            (3033666,SYM-MAT-MUL)        [3,033,666]
            (56484,MAT-SUB)              [56,484]
            (1049868,MAT-MUL)            [1,049,868]
            (1049868,MAT-MUL-ADD)        [1,049,868]
            (20682,Last-ADD)             [20,682]
        };
        \end{axis}
        \end{tikzpicture}
        \caption{KeyGen operations breakdown.}
        \label{fig:keygen_kernels}
        \end{subfigure}
        
        \begin{subfigure}[t]{0.20\textwidth}
        \centering
        \begin{tikzpicture}[baseline=(current bounding box.north)]
        \begin{axis}[
            ybar,
            bar width=5pt,
            width=\linewidth,
            height=4.6cm,
            ymin=1,
            ymax=2e10,
            ymode=log,
            log basis y=10,
            enlarge x limits={abs=5pt},
            ylabel={Clock Cycles},
            ylabel style={
                font=\tiny,
                at={(axis description cs:-0.325,0.5)},
                anchor=south
            },
            yticklabel style={font=\tiny},
            symbolic x coords={PPSPU,AU,LSS,Compare},
            xtick=data,
            x tick label style={
                font=\tiny,
                rotate=45,
                anchor=east
            },
            ymajorgrids=true,
            grid style=dashed,
            nodes near coords,
            every node near coord/.append style={
                text=purple,
                rotate=90,
                anchor=west,
                inner sep=1pt,
                font=\tiny
            },
            point meta=explicit symbolic,
            clip=false,
        ]
        \addplot[
            fill=green!15,
            draw=black,
            line width=0.4pt
        ] coordinates {
            (PPSPU,119149) [119,149]
            (AU,293197)    [293,197]
            (LSS,122007)   [122,007]
            (Compare,1)    [0]
        };
        \end{axis}
        \end{tikzpicture}
        \caption{Sign}
        \label{fig:sign_modules}
        \end{subfigure}
        \hfill
        \begin{subfigure}[t]{0.79\textwidth}
        \centering
        \begin{tikzpicture}[baseline=(current bounding box.north)]
        \begin{axis}[
            xbar,
            bar width=2.5pt,
            xmode=log,
            log basis x=10,
            xmin=10,
            xmax=1000000,
            width=\linewidth,
            height=5.3cm,
            x tick label style={font=\tiny},
            yticklabel style={font=\tiny},
            symbolic y coords={
                Copy-oil,
                Copy-vinegar,
                Copy-salt,
                VEC-M-dot-M,
                S-Sol,
                Fq-SUB,
                $\text{HASH}$,
                Gen-salt,
                Expand$_\mu$,
                LUB,
                Copy,
                C-GEN,
                EQN-GEN,
                VEC-SUB,
                VEC-MAT-MUL,
                VEC-MUL,
                SYM-MAT-MUL,
                $P_{i,2}$,
                $P_{i,1}$,
                Expand~$\text{seed}_{y}$,
                Expand$_{sk}$
            },
            ytick=data,
            point meta=explicit symbolic,
            nodes near coords={\pgfplotspointmeta},
            every node near coord/.append style={
                font=\tiny,
                text=purple,
                inner sep=1pt,
                anchor=west
            },
            xmajorgrids=true,
            grid style=dashed,
        ]
        \addplot[
            fill=green!5,
            draw=black,
            line width=0.4pt
        ] coordinates {
            (890,Expand$_{sk}$)              [890]
            (104,Expand~$\text{seed}_{y}$)   [104]
            (69552,$P_{i,1}$)                [69,552]
            (48384,$P_{i,2}$)                [48,384]
            (168642,SYM-MAT-MUL)             [168,642]
            (58482,VEC-MUL)                  [58,482]
            (58482,VEC-MAT-MUL)              [58,482]
            (1404,VEC-SUB)                   [1,404]
            (1458,EQN-GEN)                   [1,458]
            (3348,C-GEN)                     [3,348]
            (111,Copy)                       [111]
            (116314,LUB)                     [116,314]
            (71,Expand$_\mu$)                [71]
            (61,Gen-salt)                    [61]
            (87,$\text{HASH}$)               [87]
            (26,Fq-SUB)                      [26]
            (5476,S-Sol)                     [5,476]
            (1355,VEC-M-dot-M)               [1,355]
            (23,Copy-salt)                   [23]
            (60,Copy-vinegar)                [60]
            (23,Copy-oil)                    [23]
        };
        \end{axis}
        \end{tikzpicture}
        \caption{Sign operations breakdown.}
        \label{fig:sign_kernels}
        \end{subfigure}
        
        \begin{subfigure}[t]{0.20\textwidth}
        \centering
        \begin{tikzpicture}[baseline=(current bounding box.north)]
        \begin{axis}[
            ybar,
            bar width=5pt,
            width=\linewidth,
            height=3.6cm,
            ymin=1,
            ymax=2e10,
            ymode=log,
            log basis y=10,
            enlarge x limits={abs=5pt},
            ylabel={Clock Cycles},
            ylabel style={
                font=\tiny,
                at={(axis description cs:-0.325,0.5)},
                anchor=south
            },
            yticklabel style={font=\tiny},
            symbolic x coords={PPSPU,AU,LSS,Compare},
            xtick=data,
            x tick label style={
                font=\tiny,
                rotate=45,
                anchor=east
            },
            ymajorgrids=true,
            grid style=dashed,
            nodes near coords,
            every node near coord/.append style={
                text=purple,
                rotate=90,
                anchor=west,
                inner sep=1pt,
                font=\tiny
            },
            point meta=explicit symbolic,
            clip=false,
        ]
        \addplot[
            fill=green!15,
            draw=black,
            line width=0.4pt
        ] coordinates {
            (PPSPU,118094) [118,094]
            (AU,279999)    [279,999]
            (LSS,111)      [111]
            (Compare,6)    [6]
        };
        \end{axis}
        \end{tikzpicture}
        \caption{Verify}
        \label{fig:verify_modules}
        \end{subfigure}
        \hfill
        \begin{subfigure}[t]{0.79\textwidth}
        \centering
        \begin{tikzpicture}[baseline=(current bounding box.north)]
        \begin{axis}[
            xbar,
            bar width=2.5pt,
            xmode=log,
            log basis x=10,
            xmin=1,
            xmax=1000000,
            width=\linewidth,
            height=4.3cm,
            x tick label style={font=\tiny},
            yticklabel style={font=\tiny},
            symbolic y coords={
                Compare,
                Copy,
                Fq-ADD,
                VEC-M-dot-M3,
                VEC-V-dot-V2,
                VEC-M-dot-M2,
                Fq-ADD-t-plus-u,
                Fq-ADD-t-plus-t,
                VEC-V-dot-V,
                VEC-M-dot-M,
                $P_{i,2}$,
                $P_{i,1}$,
                $\text{HASH}$,
                Expand$_\mu$
            },
            ytick=data,
            point meta=explicit symbolic,
            nodes near coords={\pgfplotspointmeta},
            every node near coord/.append style={
                font=\tiny,
                text=purple,
                inner sep=1pt,
                anchor=west
            },
            xmajorgrids=true,
            grid style=dashed,
        ]
        \addplot[
            fill=green!5,
            draw=black,
            line width=0.4pt
        ] coordinates {
            (71,Expand$_\mu$)           [71]
            (87,$\text{HASH}$)          [87]
            (69552,$P_{i,1}$)           [69,552]
            (48384,$P_{i,2}$)           [48,384]
            (74520,VEC-M-dot-M)         [74,520]
            (168642,VEC-V-dot-V)        [168,642]
            (3240,Fq-ADD-t-plus-t)      [3,240]
            (3240,Fq-ADD-t-plus-u)      [3,240]
            (25434,VEC-M-dot-M2)        [25,434]
            (3348,VEC-V-dot-V2)         [3,348]
            (1512,VEC-M-dot-M3)         [1,512]
            (63,Fq-ADD)                 [63]
            (111,Copy)                  [111]
            (6,Compare)                 [6]
        };
        \end{axis}
        \end{tikzpicture}
        \caption{Verify operations breakdown.}
        \label{fig:verify_kernels}
        \end{subfigure}
        
        \vspace{-4pt}
        \caption{\malik{
        Clock cycle distribution of \textsf{QUISA} for \texttt{SL-I}.
        }
        }
        \label{fig:sl1_cycle_distribution}
    \end{figure*}

    \malik{
    To further analyze the execution cost, Fig.~\ref{fig:sl1_cycle_distribution} presents the clock cycle distribution of \textsf{QUISA} for \texttt{SL-I}. Fig.~\ref{fig:keygen_modules}, \ref{fig:sign_modules} and \ref{fig:verify_modules} report the contribution of the functional modules, while Fig.~\ref{fig:keygen_kernels}, \ref{fig:sign_kernels} and \ref{fig:verify_kernels} show the corresponding computational blocks.  
    The module-level distribution (Fig.~\ref{fig:keygen_modules}, \ref{fig:sign_modules} \& \ref{fig:sign_modules}) confirms that KeyGen is the computational bottleneck of QR-UOV, whereas Sign and Verify require substantially fewer clock cycles. It also shows that AU accounts for most of the clock cycles in all three operations. The block-level distribution (Fig.~\ref{fig:keygen_kernels}, \ref{fig:sign_kernels} \& \ref{fig:verify_kernels}) further identifies the source of this cost: SYM-MAT-MUL dominates KeyGen, while Sign is mainly dominated by SYM-MAT-MUL together with the \texttt{LUB} and \texttt{S-SoL} units. In Verify, most clock cycles are spent in the AU-based operations.
    }


    \malik{
    Table~\ref{tab:cc_sw_comparison_shake} reports the algorithmic clock-cycle counts of the proposed \textsf{QUISA} architecture and the AVX2/AVX512 QR-UOV round-2 software implementations. The software results are taken from the QR-UOV round-2 specification~\cite{qruov_round2_spec}, where the experiments were performed on an AMD EPYC~9654P processor using GCC~11.4.0 and OpenSSL~3.0.2. 
    Among the software implementations, the table shows that the AES-based \texttt{PRG} implementations require fewer algorithmic clock cycles than the corresponding SHAKE-based implementations across all QR-UOV operations and security levels.
    Since FPGA and CPU clock cycles are architecture dependent, the reported values should not be interpreted as direct execution-time or throughput comparisons. Instead, they provide a reference for the algorithmic clock-cycle requirements of the corresponding hardware and software implementations under different \texttt{PRG} choices. 
    Overall, the proposed \textsf{QUISA} architecture requires fewer algorithmic clock cycles across all QR-UOV operations and security levels than the corresponding AVX2/AVX512 software implementations. 
    }

    \begin{table*}[tb]
    \centering
    \caption{\malik{Algorithmic clock-cycle counts of the proposed \textsf{QUISA} architecture and the AVX2/AVX512 QR-UOV round-2 software implementations~\cite{qruov_round2_spec}. Green-shaded cells denote results using SHAKE-based \texttt{PRG}, whereas red-shaded cells denote results for \textsf{QUISA} with SHAKE-based \texttt{PRG} and the software implementations with AES-based \texttt{PRG}.}
    }
    \label{tab:cc_sw_comparison_shake}
    \small
    \setlength{\tabcolsep}{15.5pt}
    \begin{tabular}{l|c|c|c|c|c}
    \hline
    \multirow{1}{*}{\textbf{Security}} &
    \multirow{2}{*}{\textbf{Ref}} & 
    \multirow{1}{*}{\textbf{Parameters}} &
    \multirow{2}{*}{\textbf{Platform}} &  
    \multirow{1}{*}{\textbf{\cellcolor{green!5}Clock Cycles (M)}} &
    \multirow{1}{*}{\textbf{\cellcolor{red!5}Clock Cycles (M)}} \\  \cline{5-6}

    {\textbf{Level}} & 
    {} & 
    {\textbf{($\mathrm{q}, v, m, l$)}} & 
    {} &
    {\cellcolor{green!5}\textbf{KeyGen/Sign/Verify}} &
    {\cellcolor{red!5}\textbf{KeyGen/Sign/Verify}} \\ \hline


    \multirow{3}{*}{\texttt{SL-I}}
    & \multirow{2}{*}{\cite{qruov_round2_spec}} 
    & 127,156,54,3
    & AVX2 
    & \cellcolor{green!5}10.077/3.118/2.741
    & \cellcolor{red!5}8.587/1.586/1.224 \\   
    
    {}
    & {} 
    & 127,156,54,3
    & AVX512
    & \cellcolor{green!5}9.985/3.130/2.748 
    & \cellcolor{red!5}8.542/1.597/1.226 \\ \cline{2-6}


    {}
    & \textsf{\textbf{QUISA}} 
    & 127,156,54,3
    & \textbf{Artix-7}
    & \cellcolor{green!5}\textbf{5.330/0.535/0.399}
    & \cellcolor{red!5}\textbf{5.330/0.535/0.399} \\ \hline


    \multirow{3}{*}{\texttt{SL-III}} & 
    \multirow{2}{*}{\cite{qruov_round2_spec}} 
    & 127,228,78,3
    & AVX2 
    & \cellcolor{green!5}37.508/9.861/8.678
    & \cellcolor{red!5}{32.311/4.599/3.410} \\    

    {}
    & {} 
    & 127,228,78,3
    & AVX512
    & \cellcolor{green!5}36.717/9.824/8.604
    & \cellcolor{red!5}{31.643/4.621/3.424} \\ \cline{2-6}


    {}
    & \textsf{\textbf{QUISA}} 
    & 127,228,78,3
    & \textbf{Artix-7} 
    & \cellcolor{green!5}\textbf{22.594/1.607/1.194}
    & \cellcolor{red!5}\textbf{22.594/1.607/1.194} \\ \hline


    \multirow{3}{*}{\texttt{SL-V}} & 
    \multirow{2}{*}{\cite{qruov_round2_spec}}
    & 127,306,105,3
    & AVX2 
    & \cellcolor{green!5}109.125/24.067/20.992 
    & \cellcolor{red!5}95.579/10.504/7.463 \\

    {}
    & {} 
    & 127,306,105,3
    & AVX512 
    & \cellcolor{green!5}102.334/23.880/20.950
    & \cellcolor{red!5}88.518/10.289/7.334 \\ \cline{2-6}


    {}
    & \textsf{\textbf{QUISA}} 
    & \textbf{127,306,105,3}
    & \textbf{Artix-7} 
    & \cellcolor{green!5}\textbf{30.415/3.809/2.801}
    & \cellcolor{red!5}\textbf{30.415/3.809/2.801} \\
    
    \hline
    \end{tabular}
    \end{table*}
    

    \subsection{Area Consumption} \label{subsec:res_area}

    Table~\ref{tab:quisa_area_breakdown} reports the area breakdown of \textsf{QUISA} in terms of slices, LUTs, FFs, DSP blocks, and BRAM tiles, where no DSP slices are utilized. Since $\mathrm{q=127}$, it does not favor DSP-based multiplication in QR-UOV. 
    As expected, \texttt{PPSPU} accounts for approximately $55\%$ of the total LUT utilization of \textsf{QUISA}, driven primarily by \texttt{XOFRejSamp} and the SHAKE-128/256 wrapper that operate in parallel, contributing $\sim51\%$ and $\sim27\%$ of the \texttt{PPSPU} LUT utilization, respectively.
    The second largest block is the \texttt{LSS}, accounting for $\sim37\%$ of total LUTs and $\sim41\%$ of total FFs, comprising three submodules: \texttt{LUB} ($\sim67\%$ of \texttt{LSS} LUTs, $\sim67\%$ of \texttt{LSS} FFs) responsible for on-the-fly rank-deficient row-echelon transformation, \texttt{S-Sol} ($\sim32\%$ of \texttt{LSS} LUTs) for solution sampling, and CopyWords ($\sim1\%$) contributing negligibly.
    The \texttt{AU}, including the MAC tile ($\sim54\%$ of \texttt{AU} LUTs), is notably compact at $\sim5\%$ of total LUTs, reflecting the area efficiency of the LUT-based finite-field arithmetic.
    The \texttt{BSU}, CopyWords, and Compare modules contribute negligibly to the overall utilization, reflecting their minimal control logic.
    \texttt{Mem-2} dominates the BRAM utilization ($\sim80\%$ of total BRAM tiles) as \textsf{QUISA} computes KeyGen on-the-fly, requiring storage of the large public key matrix $\bar{P}_{i,3}$, while \texttt{Mem-1} consumes $\sim20\%$ and \texttt{Mem-3} contributes negligibly with a single BRAM tile. 
    

    \begin{table}[tb]
    \centering
    \caption{Area distribution of \textsf{QUISA} on \texttt{Artix-7} device.}
    \label{tab:quisa_area_breakdown}
    \small
    \setlength{\tabcolsep}{0.5pt}
    \begin{tabular}{l| c c c c c}
    \hline
    \multirow{2}{*}{\textbf{Design Units}} & 
    \multirow{2}{*}{\textbf{Slices}} & 
    \multirow{2}{*}{\textbf{LUTs}} & 
    \multirow{2}{*}{\textbf{FFs}} & 
    \multirow{2}{*}{\textbf{DSPs}} & 
    \textbf{BRAM} \\
    
    {} & 
    {} & 
    {} & 
    {} & 
    {} & 
    \textbf{Tiles/IPs} \\ \hline

    \texttt{PPSPU} & 5997 & 19373 & 7528 & 0 & 0 \\
    \hspace{2mm}$\lfloor$ SHAKE-128/256 wrapper & 1448 & 5181 & 3067 & 0 & 0 \\
    \hspace{4mm}$\lfloor$ KECCAK & 1396 & 5004 & 2975 & 0 & 0 \\
    \hspace{2mm}$\lfloor$ \texttt{XOFRejSamp} & 3108 & 9977 & 2929 & 0 & 0 \\
    \hspace{2mm}$\lfloor$ Output Format Block (\texttt{OFB}) & 1419 & 4172 & 1497 & 0 & 0 \\
    \hspace{2mm}$\lfloor$ Byte Stream Unit (\texttt{BSU}) & 22 & 43 & 35 & 0 & 0 \\
    Arithmetic Unit (\texttt{AU}) & 605 & 1719 & 287 & 0 & 0 \\
    \hspace{2mm}$\lfloor$ Multiplier (1 MAC) & 334 & 931 & 120 & 0 & 0 \\ 
    Linear System Solver (\texttt{LSS}) & 4543 & 13116 & 7504 & 0 & 0 \\
    \hspace{2mm}$\lfloor$ LU-Decompose (\texttt{LUB}) & 2903 & 8749 & 5018 & 0 & 0 \\
    \hspace{2mm}$\lfloor$ Sample Solution Unit (\texttt{S-Sol}) & 1555 & 4192 & 2408 & 0 & 0 \\
    \hspace{2mm}$\lfloor$ CopyWords & 85 & 175 & 78 & 0 & 0 \\
    Compare & 34 & 76 & 47 & 0 & 0 \\
    \texttt{Mem-1} ($2^{16}\times 24$) & 59 & 164 & 16 & 0 & 44/1 \\
    \texttt{Mem-2} ($2^{18}\times 24$) & 288 & 724 & 32 & 0 & 176/1 \\
    \texttt{Mem-3} ($2^{5}\times 64$) & 38 & 98 & 0 & 0 & 1/0 \\
    \hline
    Proposed \textsf{QUISA} & 11144 & 35248 & 18375 & 0 & 221/2 \\ \hline
    
    \end{tabular}
    \end{table}


    \subsection{Comparison With Related Hardware Implementations} \label{subsec:comp_hw_designs}
    \malik{
    Table~\ref{tab:fpga_comparison} compares \textsf{QUISA} with existing FPGA implementations of digital signatures. Direct comparison is inherently challenging because different schemes employ different parameter sets, leading to varying key and signature sizes that affect both hardware costs and performance. In addition, MQ-based UOV and MAYO implementations use AES for \texttt{PRG}, whereas \textsf{QUISA} relies on SHAKE, whose substantially higher computational cost directly increases the clock cycle count. Consequently, the goal of \textsf{QUISA} is not to outperform existing designs but to establish the first FPGA benchmark for QR-UOV and demonstrate a unified coprocessor that supports all three NIST security levels.
    }

    \begin{table*}[htbp]
    \small
    \caption{\malik{Comparison with the most recent SoTA digital signature schemes.}}
    \label{tab:fpga_comparison}
    \setlength{\tabcolsep}{5.0pt}
    \begin{center}
    \begin{tabular}{l|c|c|c|c|c}
    \hline
    \multirow{2}{*}{\textbf{Ref}} & 
    \multirow{2}{*}{\textbf{Scheme}} &  
    \multirow{1}{*}{\textbf{Platform}} &
    \multirow{1}{*}{\textbf{LUT/FF/DSP/}} &
    \multirow{1}{*}{\textbf{Clock Cycles}} & 
    \multirow{1}{*}{\textbf{Time ($ms$)}} \\ \cline{5-6}

    {} & {} & {\textbf{@ MHz}} & {\textbf{BRAM}} & {\textbf{(KeyGen/Sign/Verify)}} & {\textbf{(KeyGen/Sign/Verify)}} \\ \hline

    \multicolumn{6}{c}{\textbf{\malik{NIST round-3 MQ-based schemes (all uses \texttt{AES128} for \texttt{PRG} -- \textsf{QUISA} uses SHAKE)}}} \\ \hline


    \cite{HaMAYO} 
    & MAYO$_{\text{SL-I}}$
    & A7 @ 100 
    & 21k/13k/11/129
    & 996k/2M/-- 
    & 9.96/28.67/-- \\ \hline

    
    \multirow{3}{*}{\cite{whipping_MAYO}} 
    & MAYO$_{\text{SL-I}}$
    & K7 @ 100
    & 94k/32k/2/45.5  
    & 29k/95k/30k 
    & 0.30/0.96/0.30 \\ 

    {} 
    & MAYO$_{\text{SL-III}}$
    & K7 @ 100
    & 147k/52k/2/96  
    & 65k/214k/66k 
    & 0.65/2.15/0.67 \\

    {} 
    & MAYO$_{\text{SL-V}}$
    & K7 @ 100
    & 215k/73k/2/194.5  
    & 118k/384k/119k 
    & 1.19/3.89/1.19 \\ \hline

    
    \multirow{3}{*}{\cite{OV}}
    & UOV$_{\text{SL-I}}$
    & A7 @ 90
    & 28k/24k/2/66  
    & 11M/843k/284k 
    & 123.03/9.37/3.15 \\ 

    {}
    & UOV$_{\text{SL-III}}$
    & A7 @ 96
    & 38k/29k/4/184.5  
    & 16.7M/1.46M/823k 
    & 174.24/15.26/8.57 \\

    {}
    & UOV$_{\text{SL-V}}$
    & A7 @ 92
    & 77k/38k/4/359   
    & 39.0M/3.3M/1.9M 
    & 424.63/35.95/20.88 \\ \hline

     \multicolumn{6}{c}{\textbf{\malik{ML-DSA}}} \\ \hline


    \multirow{3}{*}{\cite{suparna_ml_dsa_2026}}
    & ML-DSA$_{\text{SL-II}}$
    & \multirow{3}{*}{US+ @ 230}
    & \multirow{3}{*}{54k/29k/16/29}  
    & 4.87k/24.28k/6.63k 
    & 21/106/29 (in $\mu s$) \\

    {}
    & ML-DSA$_{\text{SL-III}}$
    & {}
    & {}  
    & 8.29k/32.80k/9.73k 
    & 36/143/42 (in $\mu s$) \\

    {}
    & ML-DSA$_{\text{SL-V}}$
    & {}
    & {}  
    & 14.03k/37.59k/14.64k 
    & 61/160/64 (in $\mu s$) \\ \hline

    \multicolumn{6}{c}{\textbf{\malik{SLH-DSA (\cite{sanjay_slh_dsa_2025} uses SHA-2 for \texttt{Hash} \& results are for parameter variant-f)}}} \\ \hline


    \multirow{3}{*}{\cite{sanjay_slh_dsa_2025}}
    & SLH-DSA$_{\text{SL-I}}$
    & \multirow{1}{*}{A7 @ 110}
    & \multirow{1}{*}{6.78k/6.08k/--/8}  
    & --/5M/536k 
    & --/45.73/4.88 \\

    {}
    & SLH-DSA$_{\text{SL-III}}$
    & {A7 @ 100}
    & {15.43k/15.14k/--/12}  
    & --/9M/1.35M 
    & --/90.06/13.58 \\

    {}
    & SLH-DSA$_{\text{SL-V}}$
    & {A7 @ 100}
    & {15.51k/15.20k/--/12}  
    & --/17M/1.39M 
    & --/173.41/14.00 \\ \hline

    \multicolumn{6}{c}{\textbf{\malik{FALCON-1024 (signature verification architectures only)}}} \\ \hline


    \multirow{1}{*}{\cite{falcon_verify_2026}}
    & FALCON$_{\text{SL-V}}$
    & \multirow{1}{*}{A7 @ 200}
    & \multirow{1}{*}{8.64k/7.02k/16/3}  
    & --/--/2767 
    & --/--/13.84 (in $\mu s$) \\ \hline






    \multirow{3}{*}{\rotatebox{0}{\textsf{\textbf{QUISA}}}} 
    & \multirow{1}{*}{QR-UOV$_{\text{SL-I}}$}
    & \multirow{3}{*}{A7 @ 80}
    & \multirow{3}{*}{35k/18k/0/221}  
    & 5M/535k/399k 
    & 66.63/6.69/4.99 \\

    {} 
    & QR-UOV$_{\text{SL-III}}$
    & {}  
    & {} 
    & 22M/1.6M/1.1M 
    & 282.43/20.09/14.93 \\

    {} 
    & QR-UOV$_{\text{SL-V}}$
    & {}  
    & {} 
    & 30M/3.8M/2.8M 
    & 380.19/47.62/35.01 \\
    
    \hline
    \end{tabular}
    \end{center}    
    \end{table*}

    \malik{
    Since QR-UOV is a structured variant of UOV, the FPGA implementation in~\cite{OV} provides the most relevant comparison. At \texttt{SL-I}, \textsf{QUISA} requires $1.25\times$ more LUTs than UOV. However, this trend reverses at higher security levels, where its unified architecture maintains a nearly constant footprint of $\sim$35k LUTs, making it $2.2\times$ more LUT-efficient than UOV at \texttt{SL-V} (35k vs.\ 77k LUTs). In addition, \textsf{QUISA} eliminates DSP usage, whereas UOV requires up to four DSPs.
    For KeyGen, \textsf{QUISA} achieves $2.20\times$ and $1.30\times$ fewer clock cycles at \texttt{SL-I} and \texttt{SL-V}, respectively. At \texttt{SL-III}, however, UOV is faster because QR-UOV incurs substantially higher SHAKE-based rejection sampling overhead due to the larger increase in vinegar variables (156$\rightarrow$228 versus 160$\rightarrow$184 for UOV). Sign and Verify remain slower across all security levels due to the inherently larger QR-UOV signatures. Overall, \textsf{QUISA} demonstrates a clear area advantage while remaining competitive in KeyGen.
    }

    \malik{
     The fault-resilient MAYO implementation in~\cite{HaMAYO} is designed exclusively for \texttt{SL-I}, while \textsf{QUISA} supports all three security levels within a fixed hardware footprint, using only $1.66\times$ more LUTs. The lower Sign latency of MAYO is largely a consequence of its smaller signature size rather than the underlying hardware architecture. Since Verify clock cycles are not reported in~\cite{HaMAYO}, a complete comparison is not possible, as reflected by ``--'' in Table~\ref{tab:fpga_comparison}.
     The high-performance MAYO implementation in~\cite{whipping_MAYO} supports all three security levels by instantiating $2\times$, $3\times$, and $4\times$ round-based AES128 engines for \texttt{SL-I}, \texttt{SL-III}, and \texttt{SL-V}, respectively. Although this improves performance, the hardware cost increases with the security level, reaching $6.14\times$ more LUTs than \textsf{QUISA} at \texttt{SL-V}, together with substantially higher FF and DSP utilization. In contrast, \textsf{QUISA} offers a constant hardware footprint for all security levels without requiring DSPs. The higher BRAM utilization of \textsf{QUISA} (221 tiles) is inherent to QR-UOV, as it requires the storage of the $\bar{P}_{i,3}$ matrix during KeyGen for signature verification. 
    }
    
     \malik{Compared with standardized PQC signatures, \textsf{QUISA} implements QR-UOV using fewer LUTs (35k vs.\ 54k) than the ML-DSA design in~\cite{suparna_ml_dsa_2026}. Its higher computational cost reflects the large matrix operations and SHAKE-based rejection sampling required by QR-UOV. The SLH-DSA implementations in~\cite{sanjay_slh_dsa_2025} require fewer hardware resources, whereas \textsf{QUISA} achieves lower signing latency across all security levels. FALCON provides substantially faster verification with a smaller hardware footprint; however,~\cite{falcon_verify_2026} reports a verification-only architecture.
     }


    \section{Conclusions and Future Work}\label{sec:conclusions}
    The paper presents \textsf{QUISA}, an instruction-set-driven coprocessor architecture that implements the complete QR-UOV MQ-based signature scheme across all NIST security levels on FPGAs. 
    Our high-performance \texttt{PPSPU} module forms the core of the \textsf{QUISA} architecture. By executing SHAKE-based pseudorandom byte generation and rejection sampling in parallel, it minimizes clock cycles. To demonstrate the flexibility of the \textsf{QUISA}, we introduce a dual-datapath architecture supporting \texttt{PRG}, hash, and rejection sampling for MQ-based PQC algorithms, as well as FIPS~202-compliant \texttt{PRG} and hash operations, without additional hardware costs. All other functional modules of the \textsf{QUISA} operate iteratively. In particular, iterative memory address scheduling and shift--add folding-based modular reduction enable a compact, resource-efficient arithmetic unit. Likewise, iterative LU decomposition and sample solution modules provide a complete linear system solver with only minor clock-cycle overhead, making it applicable to other MQ-based schemes over $\mathbb{F}_{127}$.

    The parallel and iterative architectures of \textsf{QUISA} enable high performance with minimal hardware resources. For NIST \texttt{SL-V}, it performs QR-UOV KeyGen, Sign, and Verify in 380.19, 47.62, and 35.01~$ms$, respectively, using 11,144 slices, 35,248 LUTs, 18,375 FFs, and 221 BRAM tiles without DSPs on an \texttt{Artix-7} FPGA. The DSP-free design is platform-agnostic, while the ASIC implementation requires only replacing the BRAMs (\texttt{MEM-1} and \texttt{MEM-2}) with SRAM macros generated by commercial memory compilers.

    \section*{Acknowledgments}
    The authors would like to thank the QR-UOV team for their valuable discussions and continued support throughout this work. We also thank Ryan Bevin (Queen's University Belfast, Northern Ireland, United Kingdom) for his assistance in setting up the reference implementation and Muhammad Rashid (Umm Al-Qurrah University, Makkah, Saudi Arabia) for his insightful technical comments and careful proofreading of the manuscript.

    \bibliographystyle{IEEEtran}
    \bibliography{references}
    \input{biography}

\vfill

\end{document}

%% file: biography.tex
\vspace{-15mm}

\begin{IEEEbiography}[{\includegraphics[width=1in,height=1.25in,clip,keepaspectratio]{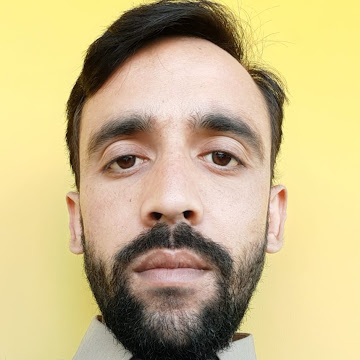}}]{Malik Imran} 
        is a Research Fellow at the Centre for Secure Information Technologies (CSIT), Queen’s University Belfast, Northern Ireland. He received his Ph.D. in Information and Communication Technologies from Tallinn University of Technology (TalTech), Estonia, in 2023. He received M.S. and B.S. degrees in Telecommunication and Networks and Computer Engineering from Pakistan in 2015 and 2011, respectively. His research interests focus on secure-by-design hardware accelerators for cryptographic primitives, with emphasis on post-quantum cryptography, homomorphic encryption and privacy computing. 
    \end{IEEEbiography}

    \vspace{-10mm}

    \begin{IEEEbiography}[{\includegraphics[width=1in,height=1.25in,clip,keepaspectratio]{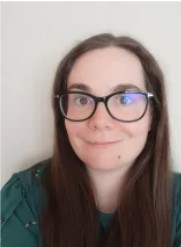}}]{Ciara Rafferty} 
        (M'14, SM'24) is a Senior Lecturer at the Centre for Secure Information Technologies (CSIT) in Queen’s University Belfast, Northern Ireland. She received her BSc degree in Mathematics with Extended Studies in Germany from Queen's University Belfast in 2011 and the PhD degree in Electrical and Electronic Engineering from Queen's University Belfast in data security in 2015. Her research interests include post-quantum cryptography,  privacy enhancing technologies and optimised hardware architectures of cryptographic primitives.
    \end{IEEEbiography}

    \vspace{-10mm}

    \begin{IEEEbiography}[{\includegraphics[width=1in,height=1.25in,clip,keepaspectratio]{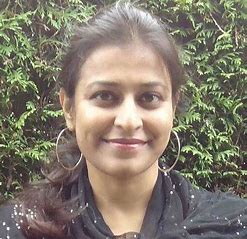}}]{Ayesha Khalid} 
        (M'2017--SM'2022) is a Senior Lecturer with Queen's University of Belfast, Belfast, U.K. She received her the B.E. degree in Computer Systems Engineering from the National University of Sciences and Technology (NUST), Pakistan, and the M.S. degree in Electrical Engineering from the Center for Advanced Studies in Engineering (CASE), University of Engineering and Technology, UET-Taxila. Her research interests include lattice-based cryptography, embedded systems security, side channel attacks, and cryptographic hardware. She was a recipient of the DAAD Scholarship Award for her Ph.D. studies at RWTH Aachen, Germany.
    \end{IEEEbiography}

    \vspace{-10mm}

    \begin{IEEEbiography}[{\includegraphics[width=1in,height=1.25in,clip,keepaspectratio]{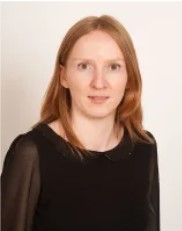}}]{Maire O'Neill} 
        (M’03-SM’11) received the M.Eng. degree with distinction and the PhD degree in electrical and electronic engineering from Queen’s University Belfast, U.K., in 1999 and 2002, respectively. She is a Chair of Information Security at Queen’s and holds an EPSRC Leadership fellowship to conduct research into next generation data security architectures. She previously held a UK Royal Academy of Engineering research fellowship from 2003 to 2008. She has authored two research books and has more than 100 peer-reviewed conference and journal publications. Her research interests include hardware cryptographic architectures, side channel analysis, physical unclonable functions and post-quantum cryptography. In 2014 she was awarded a Royal Academy of Engineering Silver Medal, which recognises outstanding personal contribution by an early or mid-career engineer that has resulted in successful market exploitation.
    \end{IEEEbiography}